\documentclass[preprint2]{aastex}

\shorttitle{Rotational periods of very young brown dwarfs
and very low-mass stars in Cha\,I}
\shortauthors{Joergens et al.}

\begin{document}


\title{Rotational periods of very young brown dwarfs \\
and very low-mass stars in Cha\,I\altaffilmark{1}}


\author{V. Joergens\altaffilmark{2}, M. Fern\'andez\altaffilmark{3},
J. M. Carpenter\altaffilmark{4} and R. Neuh\"auser\altaffilmark{2,5}}


\altaffiltext{1}{Based on observations obtained at the European Southern
	Observatory at La Silla in program 65.L-0629.}

\altaffiltext{2}{Max-Planck-Institut f\"ur Extraterrestrische Physik,
              Giessenbachstrasse 1, D-85748 Garching, Germany. Email: 
	      viki@mpe.mpg.de}

\altaffiltext{3}{Instituto de Astrof\'{\i}sica de Andaluc\'{\i}a
              (CSIC), Apdo. 3004, E-18080 Granada, Spain. 
	      Email: matilde@iaa.es}

\altaffiltext{4}{Department of Astronomy, MS 105-24, 
		California Institute of Technology, 
		1201 East California Boulevard, Pasadena, CA 91125.
		Email: jmc@astro.caltech.edu}

\altaffiltext{5}{Astrophysikalisches Institut der Universit\"at von Jena,
		Schillerg\"asschen 2-3, D-07745 Jena, Germany.
		Email: rne@astro.uni-jena.de}


\begin{abstract}
We have studied the photometric variability of very young
brown dwarfs and very low-mass stars (masses well below 
0.2\,M$_{\odot}$) in the Cha\,I star forming region. 
We have determined photometric periods in the Gunn i and R band for the 
three M6.5--M7 type brown dwarf candidates 
Cha\,H$\alpha$\,2, Cha\,H$\alpha$\,3 and
Cha\,H$\alpha$\,6 of 2.2 to 3.4 days.
These are the longest photometric periods found for any brown 
dwarf so far. If interpreted as rotationally induced they 
correspond to moderately fast rotational velocities, which is fully 
consistent with their $v \sin i$ values 
and their relatively large radii.
We have also determined periods for the two M5--M5.5 type
very low-mass stars B\,34 and CHXR\,78C.
In addition to the Gunn i and R band data, we have 
analysed JHK$_\mathrm{S}$ monitoring data of the targets, which have been
taken a few weeks earlier and confirm the periods found in the optical data.
Upper limits for the errors in the period determination are between 
2 and 9 hours.
The observed periodic variations of the brown dwarf candidates as 
well as of the T~Tauri stars are interpreted as modulation of the 
flux at the rotation period by magnetically driven surface 
features, on the basis of a consistency with $v \sin i$ values as 
well as (R-i) color variations typical for spots. 
Furthermore, the temperatures even for the brown dwarfs in the sample 
are relatively high ($>$\,2800\,K) because the objects are very 
young. Therefore,
the atmospheric gas should be sufficiently ionized for the 
formation of spots on one hand and the temperatures are
too high for significant dust condensation and hence variabilities due
to clouds on the other hand.
A comparison with rotational properties of older brown dwarfs 
shows that most of the acceleration of brown dwarfs
takes place within the first 30\,Myr or less.
If magnetic braking plays a role this suggests that
the disk dissipation for brown dwarfs occurs between a few Myrs
and 36\,Myr.
\end{abstract}




\section{Introduction}
A photometric monitoring campaign of bona fide and
candidate brown dwarfs in the Cha\,I
star forming cloud has been carried out in two filters
in order to study the time dependence of their brightness and color.
It is known that magnetically driven surface features (spots) of stars
modulate the brightness of the star as it rotates
(e.g. Bouvier et al. 1993 and references therein).
The presented photometric observations of brown dwarfs are aimed at the study of 
such spot-driven variabilities in the substellar regime in order to test if
brown dwarfs have magnetic spots. Furthermore, since surface features
modulate the emitted flux \emph{at the rotational period}, the photometric study
is aimed at the determination of rotational periods for brown dwarfs.

Rotational periods are fundamental (sub)stellar properties.
The knowledge of rotational periods for objects covering a wide range of
ages is important for the understanding of the evolution of angular momentum.
Scanning the period--age--diagram observationally for brown dwarfs with ages
of less than 100\,Myr provides in addition a test of substellar 
evolutionary theories since the onset of deuterium burning is expected to have
an observable effect on the angular momentum evolution.
The contraction of brown dwarfs is expected to be temporarily decelerated or
even stopped in the first several million years of their lifetime due to the 
ignition of deuterium (e.g. Burrows et al. 2001).
However, since age estimates are also often model dependent, the significance of 
such a test might be somewhat limited.
Furthermore, rotation rates are critical parameters for 
rotationally induced phenomena, like dynamo activity (supposed to cause surface
spots) and meteorological processes.

For solar-type main sequence stars there is a correlation
between rotation and activity: 
the faster the rotation the more active the star is, measured 
in terms of chromospheric H$\alpha$ and Ca\,II emission, 
flare activity as well as coronal X-ray emission. 
There are indications that this relation becomes invalid for later
spectral types (late M) and lower masses. In particular near and below 
the substellar limit several rapid rotators are found with no or very little
signs of chromospheric activity 
(e.g. Basri \& Marcy 1995; Delfosse et al. 1998; Gizis et al. 2000).

For very young stars on the pre-main sequence (T~Tauri stars)
the relation between 
activity and rotation is still a matter of debate:
correlations between rotation and X-ray emission have been found for 
T~Tauri stars in Taurus 
(Bouvier 1990; Neuh\"auser et al. 1995; Stelzer \& Neuh\"auser 2001),
whereas a recent publication by Feigelson et al. (2002) reports the absence of
a connection between X-ray emission and rotation for a large sample of 
T~Tauri stars in Orion.
H$\alpha$ emission on the other hand is not a definite activity indicator 
for very young objects, because they often have circumstellar accretion 
disks, which are significant additional H$\alpha$ emission sources. 
A study of the rotation-activity-relation in the 
substellar regime at this very young age is hampered up to now by the 
lack of observational constraints of rotation parameters:
$v \sin i$ values are known for two very young brown dwarfs and twelve
brown dwarf candidates (Joergens \& Guenther 2001, White \& Basri 2003).
Furthermore, Bailer-Jones \& Mundt (2001) found photometric periods
for two very young late-M dwarfs in $\sigma$\,Ori, which might be 
rotational periods.

For very cool objects (spectral type late M, L) 
surface spots might not play a significant role
but another important process
may affect the time dependence of the observed flux of the objects:
below a temperature of about 2800\,K the condensation of dust sets in
(e.g. Tsuji et al. 1996a,b; Allard et al. 1997; Burrows \& Sharp 1999).
Inhomogeneities in dust cloud coverages may cause
observable photometric variations.

In spite of the demonstrated significance of rotational periods, 
the number of brown dwarfs with known rotational periods is rather small
(see Sect.\,\ref{hitherto}). This is particularly the case 
at very young ages.
We have therefore carried out a photometric monitoring 
campaign of brown dwarfs and very low-mass stars in the Cha\,I
star forming cloud and determined periods for three brown
dwarf candidates.
The targets are introduced in Sect.\,\ref{sample}, the data acquisition and analysis 
is described in Sect.\,\ref{phot} and \ref{timeseries}. 
The results are presented in Sect.\,\ref{results} 
and discussed in Sect.\,\ref{discussion}, followed by a summary
in Sect.\,\ref{concl}.

\section{Hitherto known periods for brown dwarfs}
\label{hitherto}

Several studies of the photometric behavior of late-M and L-dwarfs
have been carried out so far. 
They have led to the detections of a handful of brown dwarfs showing periodic 
variabilities and of many, 
mostly substellar L-dwarfs, showing non-periodic variabilities.
We compiled all photometric periods that we have found for brown dwarfs
in the literature in Table\,\ref{literature} and ordered them by 
increasing age.

The nature of the detected periodic variations are not finally clarified.
Tinney \& Tolley (1999) report variations of an M9 dwarf in 
narrow band filters, which are sensitive 
to changes in TiO absorption features and therefore to clouds.
%
Bailer-Jones \& Mundt (2001) detected significant periods
below one day for four late-M and L dwarfs, 
among them the very young object S\,Ori\,31. Furthermore, they have
found hints for a period for S\,Ori\,33.
Out of these five periods four might be rotational periods according to the authors.
The detected periods below one day for S\,Ori\,31 and S\,Ori\,33
imply very rapid rotational velocities of the order of 100\,km\,s$^{-1}$
(cf. Sect.\,\ref{radii}) if confirmed as rotational periods.
Furthermore, they find non-periodic variations for several L-dwarfs
with time scales of hours and
suggest that one sees the formation and dissipation of clouds
rather than rotationally driven features. 
Mart\'\i n, Zapatero Osorio \& Lehto (2001) report a varying periodicity
found in I band photometry of the M9.5 dwarf BRI\,0021-0214.
The authors suggest that the variations are caused by inhomogeneous
clouds rather than spots because the object is very inactive in terms of
H$\alpha$ and X-ray emission 
despite a very fast rotation.
Clarke et al. (2002) found periodic variations of the brightness of Kelu-1.
The detected period of 1.8\,h is consistent with the rotational velocity. 
Nevertheless, Clarke \& Tinney (2002) found no evidence for variability
in dust sensitive molecular lines and the nature of the variations 
remains still unclear.
Recently, Gelino et al. (2002) found photometric variabilities for 
several L-dwarfs.
These features show no periodicities or at least no persistent periodicities
and are therefore rather caused by rapid evolution of atmospheric features
than being rotationally induced.
Eisl\"offel \& Scholz (2001) studied the photometric behavior 
of very low mass stars and brown dwarfs in the young ($\sim$36\,Myr) 
cluster IC4665 and report the finding of rotational periods 
below one day for five candidate brown dwarfs. 
Details on periods and amplitudes will be given in a 
forthcoming publication (pers. comm.).

\section{Sample}
\label{sample}

The targets of our observations are twelve 
low-mass M6--M8--type objects,
Cha\,H$\alpha$\,1 to 12, located in the center of the Cha\,I star forming
cloud with an age of 1--5\,Myr (Comer\'on, Rieke \& Neuh\"auser 1999; 
Comer\'on, Neuh\"auser \& Kaas 2000).
Their masses are below or near the border line separating brown dwarfs
and very low-mass stars. 
Four of them are bona fide brown dwarfs 
(Neuh\"auser \& Comer\'on 1998, 1999; Comer\'on et al. 2000).
We like to note, that all three brown dwarf candidates, 
for which we are presenting rotational periods in this paper, 
Cha\,H$\alpha$\,2, 3 and 6, have masses below the hydrogen 
burning mass limit but they are \emph{candidates}
because the error bars of their effective temperatures extend 
into the stellar regime (cf. Comer\'on et al. 2000).

A first study of their rotational properties
has been carried out 
by Joergens \& Guenther (2001), who measured the rotational
broadening of spectral features in high-resolution UVES spectra.
They found $v \sin i$ values within the range of 8 to 26\,km\,s$^{-1}$
for nine out of the twelve objects. 


Furthermore, the very low-mass T~Tauri stars B\,34, CHXR73 and CHXR\,78C,
in the same field-of-view, have been studied.

\section{Photometry}
\label{phot}
We monitored a 13$^{\prime} \times 13^{\prime}$ region in the Cha\,I cloud 
photometrically in six consecutive half
nights with DFOSC at the Danish 1.5\,m telescope at ESO, La Silla, 
Chile. Images have been obtained in the Bessel R and the Gunn i filter
between 2000 May 31 and June 5.
The first two nights have been partly cloudy and therefore fewer
images (with lower S/N) have been taken in these two nights.

The exposure times have been chosen to detect periods on time scales
of expected rotational periods of the objects. Projected rotational
velocities $v \sin i$ (Joergens \& Guenther 2001)
indicate that their rotational periods are within the range of a few days.
The objects span a large dynamical range (I=13.6 to 17.4\,mag)
therefore we obtained Gunn i band images with two different exposure times
of about 400\,s and about 900\,s. R band images have been taken with only
one exposure time of about 1000\,s.

We performed aperture photometry for the twelve bona fide and candidate brown 
dwarfs Cha\,H$\alpha$\,1 to 12, the very low-mass T~Tauri stars 
CHXR\,73, CHXR\,78C, B\,34 as well as for several
reference stars in the field with 
IRAF\footnote{IRAF is distributed by the National Optical
   Astronomy Observatories,
   which is operated by the Association of Universities for Research in
   Astronomy, Inc. (AURA) under cooperative agreement with the National
   Science Foundation.}. 
The sky background was determined from a source-free
annulus and subtracted from the object counts.
The T~Tauri stars CHXR\,74 and Sz\,23 are also in the field of view
but have been saturated.
The calculation of differential magnitudes allowed us to compensate 
for variable atmospheric conditions at least to a certain degree.
Three reference stars have been chosen carefully for each
filter based on the criteria of 
constant brightness over the time of observations, good S/N as well as
similar brightness as the targets. 
From the analysis of these
reference stars we estimate the photometric error to be about 
0.015\,mag or less:
the standard deviation of the reference stars
is 0.009 to 0.014\,mag in the Gunn i band and 0.005 to 0.006\,mag in the R band
(cf. Fig.\,\ref{b34_lc} and Fig.\,\ref{78c_lc} to \ref{cha6_lc} for the 
dispersion of the reference stars).

\section{Time series analysis}
\label{timeseries}
We applied the string-length method (Dworetsky 1983) in order to
search for periodicity in the obtained light curves of the targets.
This method is ideally suited for unevenly spaced data as in our case. 
The algorithm phase folds the data with a trial period and calculates
the string length between successive data points. This is done for all periods
within a given period range. The period which generates the minimum
string length is the most likely period.

The significance of the detected periods was estimated by cross checking
a randomized data set sampled with the same time steps as 
the real data but with arbitrary magnitudes within
the limits of the real magnitudes. 
For each suspected period, we checked 10000 randomized data sets.
The percentage of samples that have a longer (i.e. less significant) 
string length for any period than that of the suspected
period yields the confidence level (a 99.99\% confidence level
corresponds to the case that all randomized samples have a
longer string length than the string length for the suspected period).
Besides this significance test, a major emphasis was put on a direct
check by eye of the original as well as the phase folded light curves.

It is well known that in addition to intrinsic periodicities of the 
monitored objects the light curves may show alias periods 
due to the sampling rate of the data.
The most common alias periods $P_\mathrm{false}$ are
\begin{equation}
\label{alias}
P_\mathrm{false}^{-1} = 1.0027\,\mathrm{d}^{-1} \pm P_\mathrm{true}^{-1}
\end{equation}
(Dworetsky 1983). This equation relates the true period with 
possible alias periods, which are inherent to the data due to an 
observing frequency of one sidereal day.

In general, more images have been obtained in the Gunn i band than in the 
R band and in addition the Gunn i band data have a higher S/N.
Therefore for each object, firstly the Gunn i band data were studied and
it was given them a higher weight.
We have searched for periods in the range of 1.5 hours to 5 days.
The minimum period is set by twice the average sampling rate of 
about 45\,min. The maximum is chosen to be slightly smaller than 
the total time coverage of about 5.5\,d.

\section{Results}
\label{results}

We have measured significant periods for three brown dwarf candidates
(Cha\,H$\alpha$\,2, 3, 6) as well as for two very low-mass T~Tauri stars 
(B\,34, CHXR\,78C) within the range of 2.2\,d to 4.8\,d
and with Gunn i band amplitudes
between 0.06\,mag and 0.14\,mag (Table\,\ref{largetable} and 
Fig.\,\ref{b34_lc}, \ref{78c_lc}--\ref{cha6_lc}).

For Cha\,H$\alpha$\,4, 5, 8, 12 and CHXR\,73 no clear variations are detected
and limits for variability amplitudes for these objects are derived
(Table\,\ref{largetable}). 
The S/N of Cha\,H$\alpha$\,1, 7, 9, 10 and 11 in the obtained images
is too low for a further exploitation.
These objects should be re-observed with a larger telescope.

Details of the results of the period analysis on individual objects as 
well as original 
and phase folded light curves are presented in the following subsections.
Objects are ordered by decreasing mass: from 
about 0.12\,M$_{\odot}$ for B\,34 to about 0.05\,M$_{\odot}$ for 
Cha\,H$\alpha$\,6. These masses are rough estimates based on  
a comparison with theoretical evolutionary tracks by Baraffe et al. 
(1998) for the Cha\,H$\alpha$ objects (Comer\'on et al. 2000) and
by Burrows et al. (1997) for the T~Tauri stars B\,34 and CHXR\,78C
(Comer\'on et al. 1999), respectively. 

Recently, a near-infrared (JHK$_\mathrm{S}$) photometric monitoring campaign
of the Cha\,I cloud has been carried out 
(Carpenter et al. 2002), which includes all the targets of our
sample. These observations contain between 11 and 25 points per target 
obtained in 10--12
separate nights, mostly in 2000 April and May and for some targets
(B\,34, CHXR\,78C and Cha\,H$\alpha$\,3) there is an additional 
point in 2000 January.

After checking that the period analysis performed for our data clearly excludes
periods shorter than one day, we have decided to take the data of 
Carpenter et al. (2002) into account for our analysis. 
Our Gunn i and R band light curves are well-sampled but
have a total time basis of less than six days and are therefore hardly providing 
strong support for periods longer than three days. 
On the other hand, the JHK$_\mathrm{S}$ data include just
one or two points per night but cover several weeks. 
Therefore, they are
the ideal complement to our data set and increase the total time basis to 
55--59 days 
(not taking into account the single measurement from 2000 January). 

Since the two data sets correspond to different filters, we have joined
only data taken with the Gunn i and J filters, for which the difference in
effective wavelength is smaller. In order to deal with the
differences in amplitude and average brightness of the objects in the two 
different bands, we have normalized each
data set by subtracting its mean and dividing by its standard
deviation\footnote{If normalization is done by subtracting the middle value of
the light curve and dividing by its amplitude, the same results are obtained
but, mathematically, the first method is more consistent.}. 
Thereafter, we have carried out a period analysis of the objects
in the so-created joined Gunn i and J data set. The results
are given for each object at the end of the corresponding 
subsection. 
Upper limits for the error of the period determinations have been
computed from Nyquist's frequency
based on a time basis of 55 to 59 days. 
They range from 9 to 2 hours.
The error bars of the J band data, except for B\,34, 
are between a factor 0.3 and 0.4 of the amplitude observed in that
band.
For B\,34 the error bars are smaller, namely between a factor 0.15 and 0.23
of its J amplitude.

Only for the plots, the normalized iJ light curves have been 
additionally divided by their amplitudes for clarity.

\subsection{B\,34 ($\sim$ 0.12\,M$_{\odot}$)}

\begin{figure}[h]
\begin{center}
\includegraphics[width=.5\textwidth]{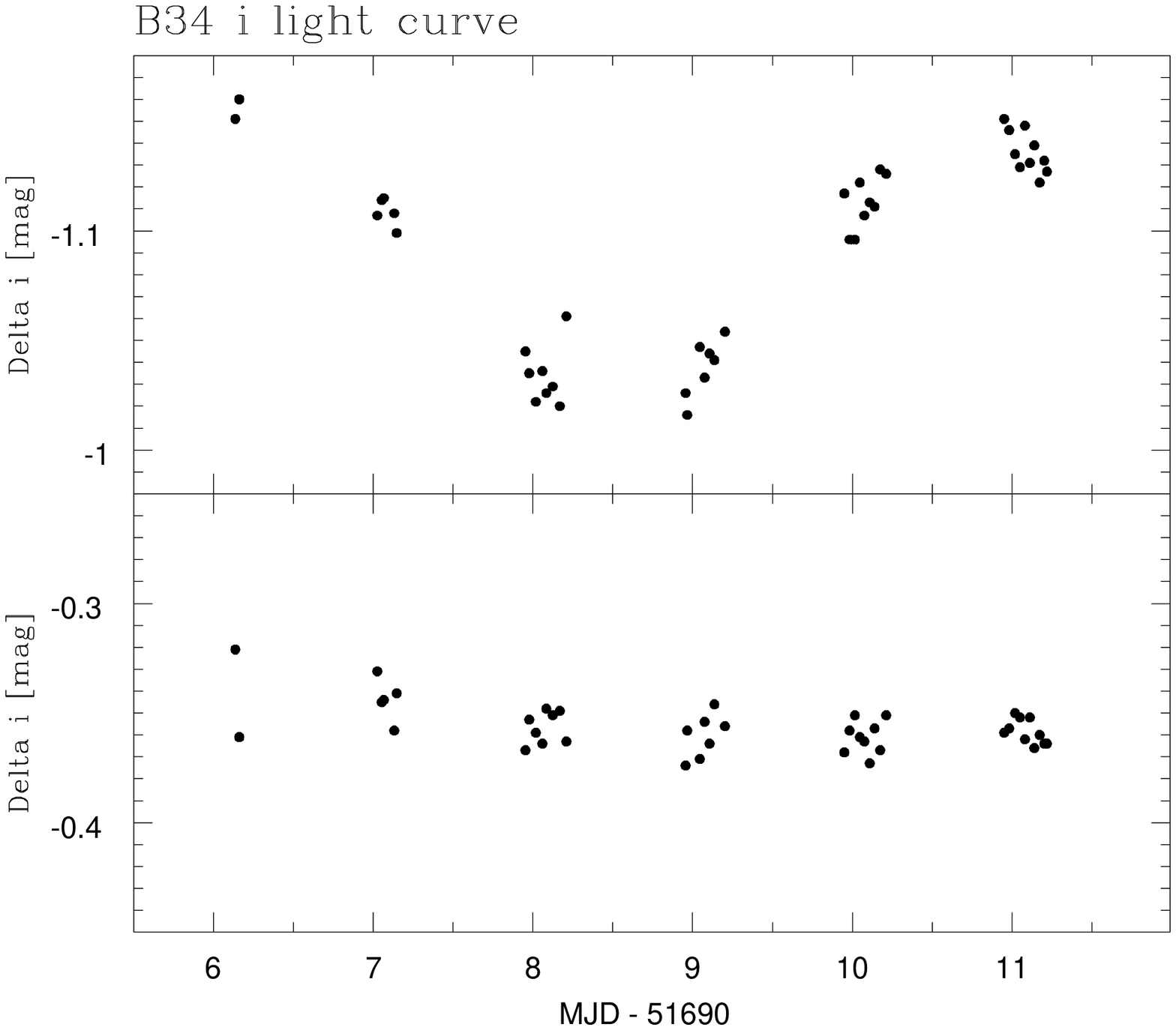}
\includegraphics[width=.5\textwidth]{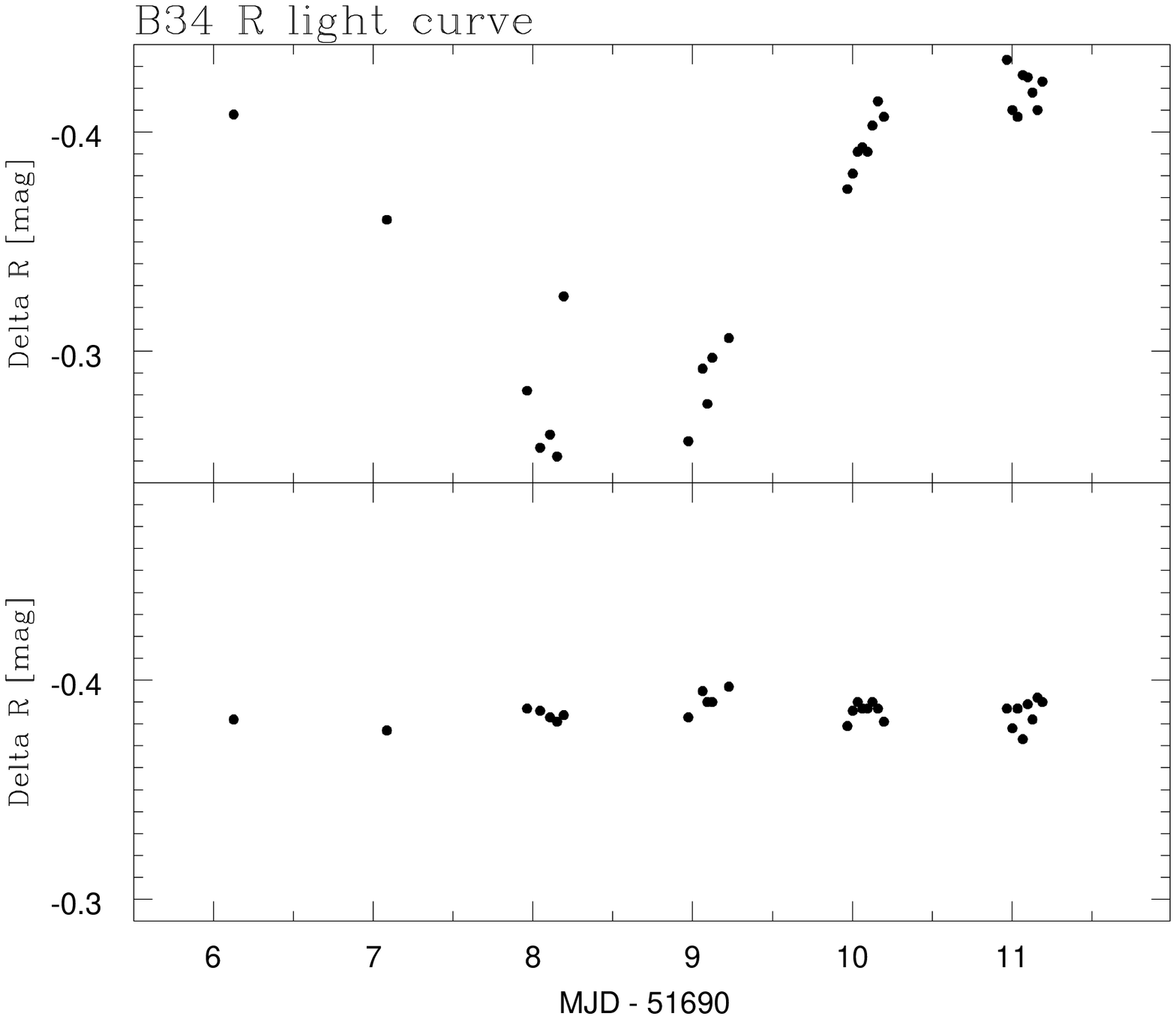}
\includegraphics[width=.5\textwidth]{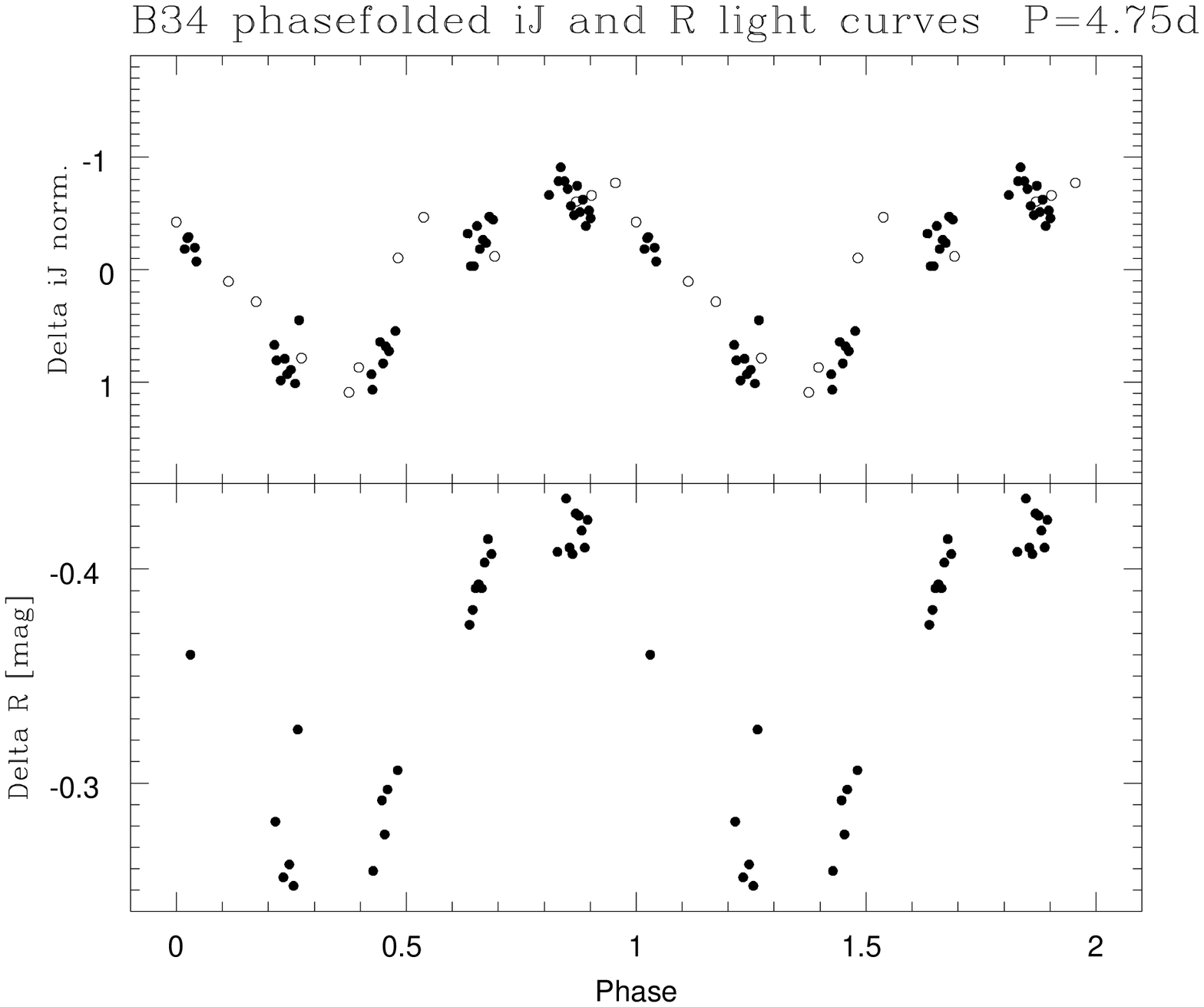}
\end{center}
\caption[]{\label{b34_lc} Light curves 
of the very low-mass T~Tauri star B\,34.
Top and middle panel: Gunn i and R band original light 
curves (relative magnitudes), each with reference star magnitudes plotted below
for comparison. Bottom panel: joined iJ and R light curves phase 
folded with the determined period. 
Filled circles: our Gunn i and R data.
Open circles: J band data from Carpenter et al. (2002).
}
\end{figure}

The brightness of the T~Tauri star B\,34 varies with a peak-to-peak 
amplitude of 0.14\,mag 
in the Gunn i band and 0.18\,mag in the R band during the 
six nights of our observations (Fig.\,\ref{b34_lc}). 
The light curves show a smooth variability
apart from one runaway data point in the 3rd night, 
which is present in both filters 
and shows a larger deviation in the R band than in the Gunn i band,
which is consistent with an intrinsic increase of the brightness due 
to a flare-like event. 

The period search analysis of the Gunn i light curve of B\,34 yields
a clear period of 4.5\,d with a confidence level above 99.99\%. 
Searching for periods in the R band data yields rather a plateau 
with a range of 4.2\,d to 4.9\,d
for the string length than a clear minimum, i.e. the period search algorithm
is unable to distinguish between the significance of periods within this range.
This can be understood if one takes into account that 
these periods are very close to the total time basis of the data set 
($\sim$5.5\,d) and that the R band brightness of the first two nights 
is only constrained by one data point each. 
Therefore, any errors of these points greatly affect the period search.

\begin{figure}[h]
\begin{center}
\includegraphics[width=.5\textwidth]{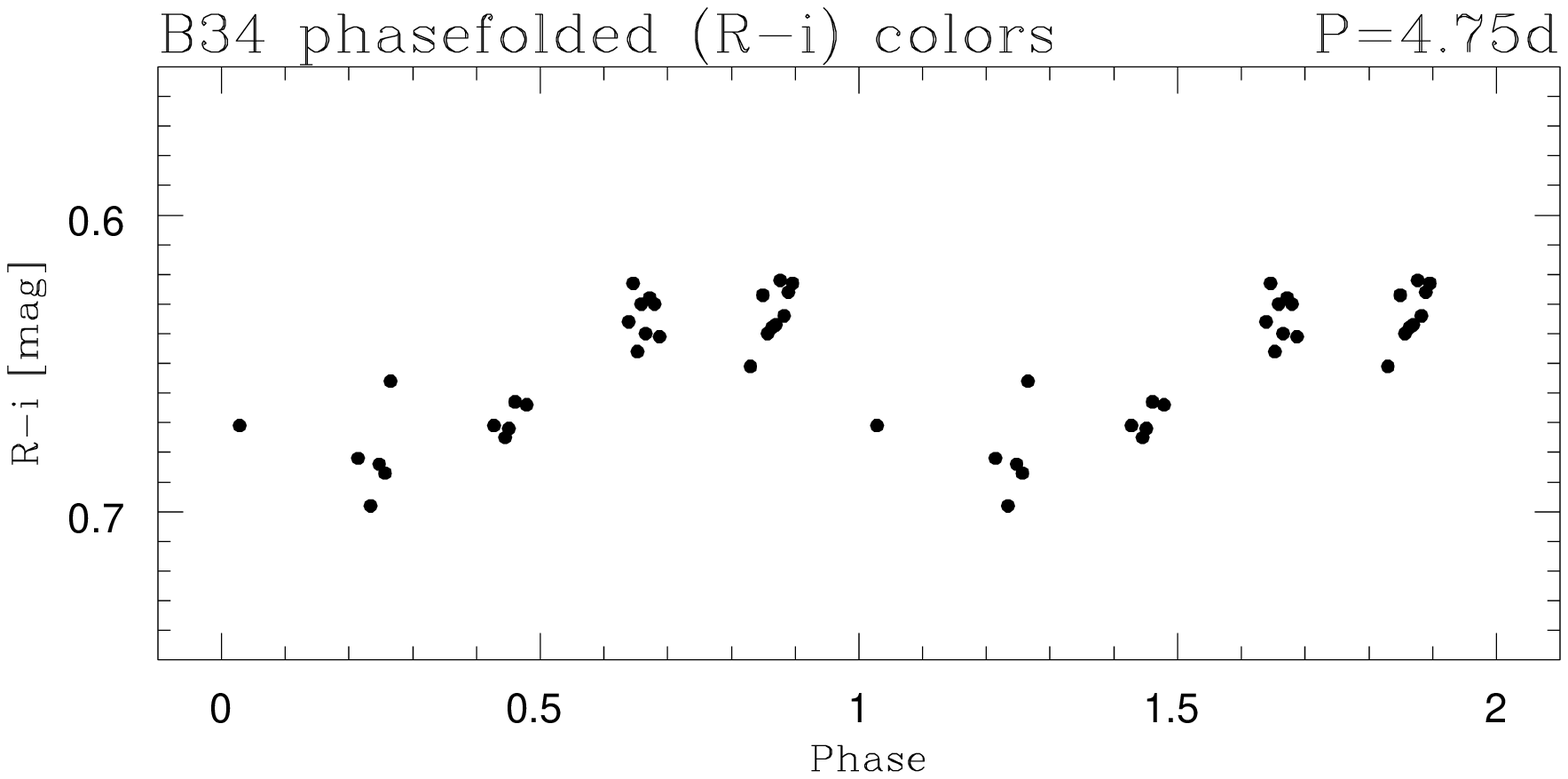}
\caption[]{\label{b34_Rmini}
Phase folded (R-i) colors of B\,34.
}
\end{center}
\end{figure}

The period analysis carried out on the Gunn i and J joined data set
solves the uncertainty of the mentioned plateau and gives 4.75$\pm$0.38\,d
as the best period.
The data folded with that period are shown in Fig.\,\ref{b34_lc} (bottom).
The phase folded (R-i) color curve (Fig.\,\ref{b34_Rmini}) shows that B\,34
is redder during minimum light as expected for brightness variations caused 
by star spots.

\subsection{CHXR\,78C ($\sim$ 0.09\,M$_{\odot}$)}
\begin{figure}[h]
\begin{center}
\includegraphics[width=.5\textwidth]{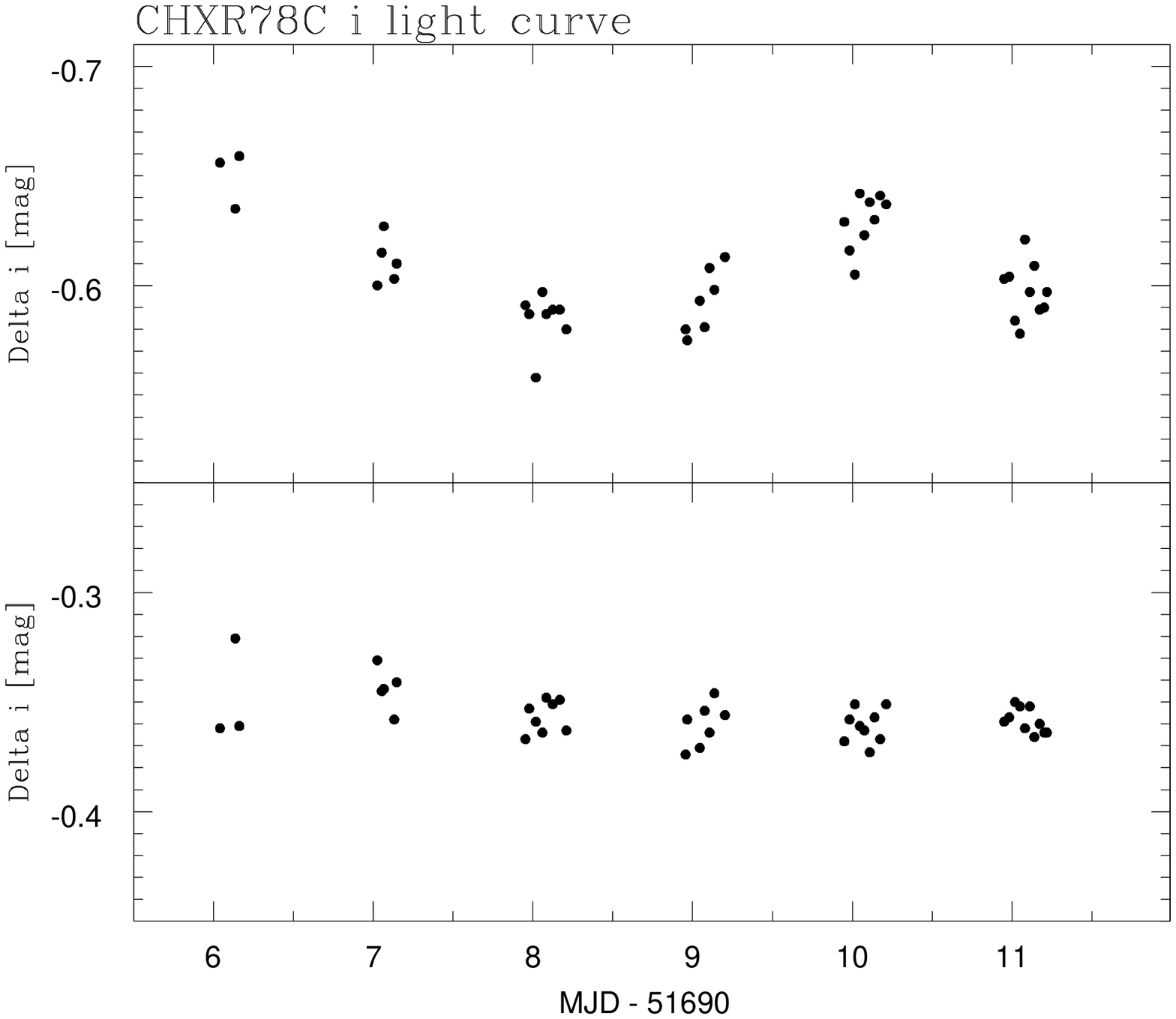}
\includegraphics[width=.5\textwidth]{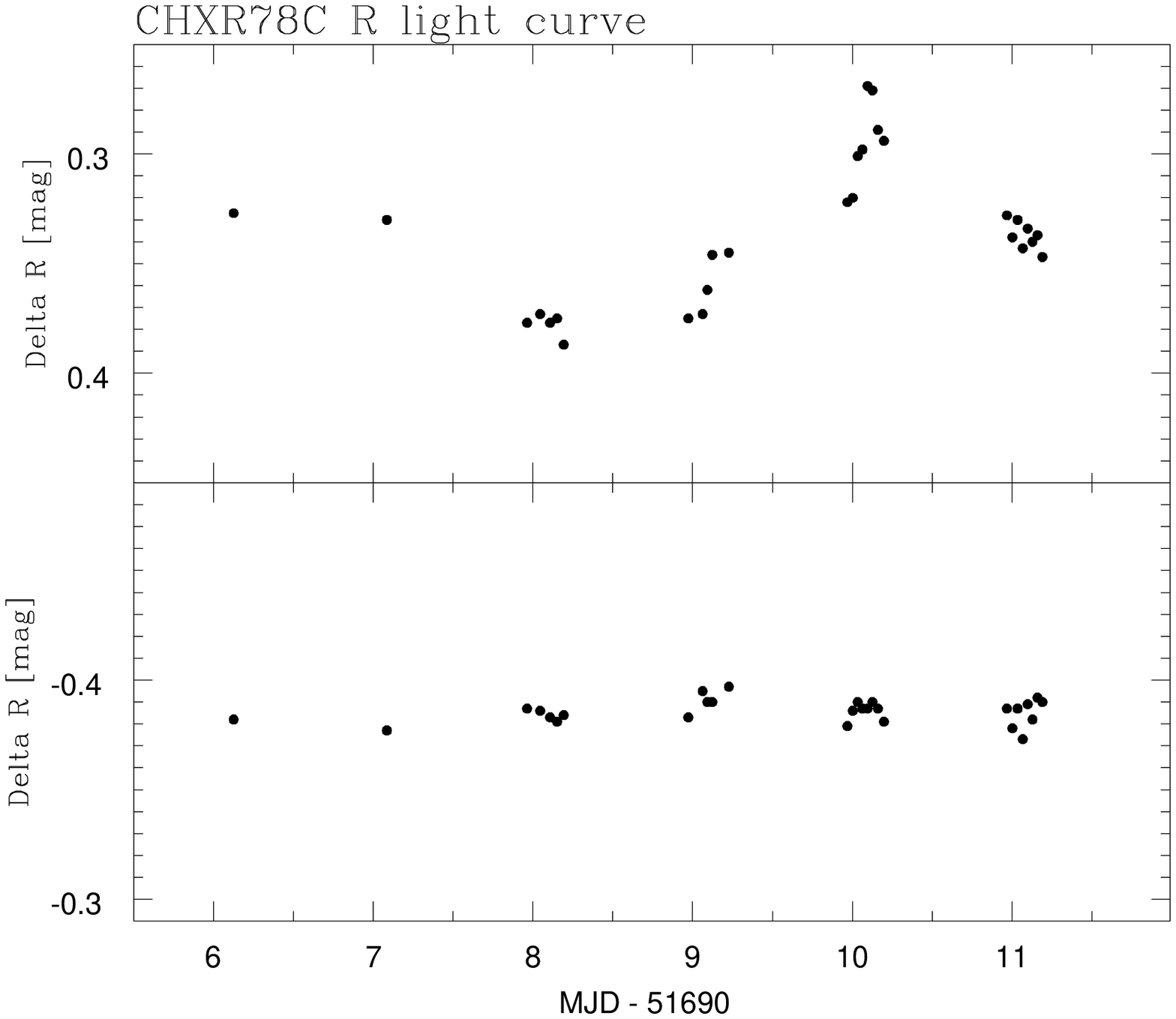}
\includegraphics[width=.5\textwidth]{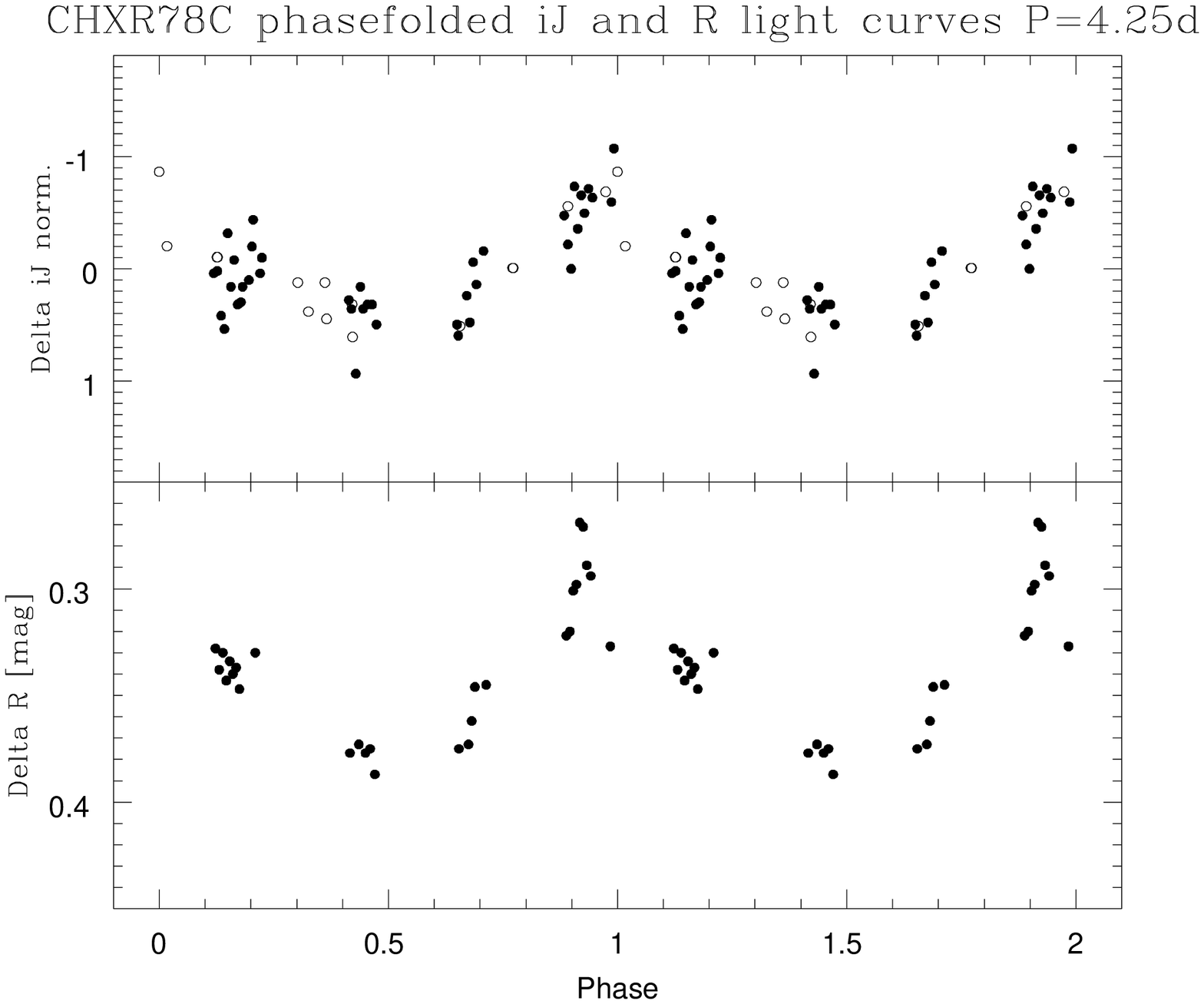}
\end{center}
\caption[]{\label{78c_lc} Light curves 
of the very low-mass T~Tauri star CHXR\,78C.
Panels are as in Fig.\,\ref{b34_lc}.
}
\end{figure}

The T~Tauri star CHXR\,78C changes its brightness with an amplitude 
of 0.07\,mag in Gunn i and 0.10\,mag in R (Fig.\,\ref{78c_lc}).
The period search algorithm returns a period of 4.0\,d for the 
Gunn i band light curve and a period of 3.9\,d for the R band light curve, 
both with a confidence level above 99.99\%.

The period analysis carried out on the Gunn i and J joined data set
yields a slightly longer period of 4.25$\pm$0.31\,d. 
The folded Gunn i and R band data confirm it as the best period
(Fig.\,\ref{78c_lc}, bottom).

\subsection{Cha\,H$\alpha$\,2 ($\sim$ 0.07\,M$_{\odot}$)}

\begin{figure}[t]
\begin{center}
\includegraphics[width=.5\textwidth]{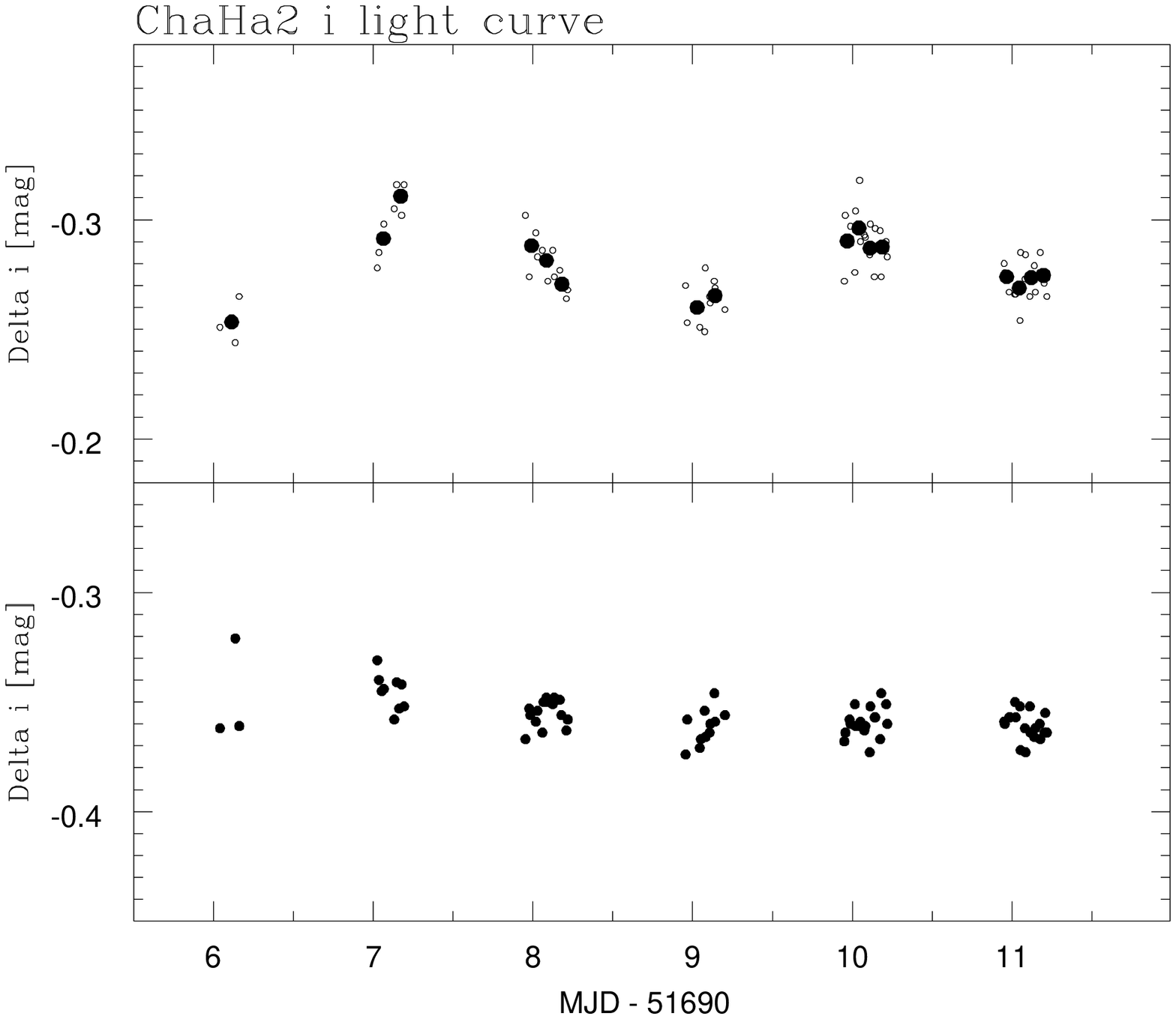}
\includegraphics[width=.5\textwidth]{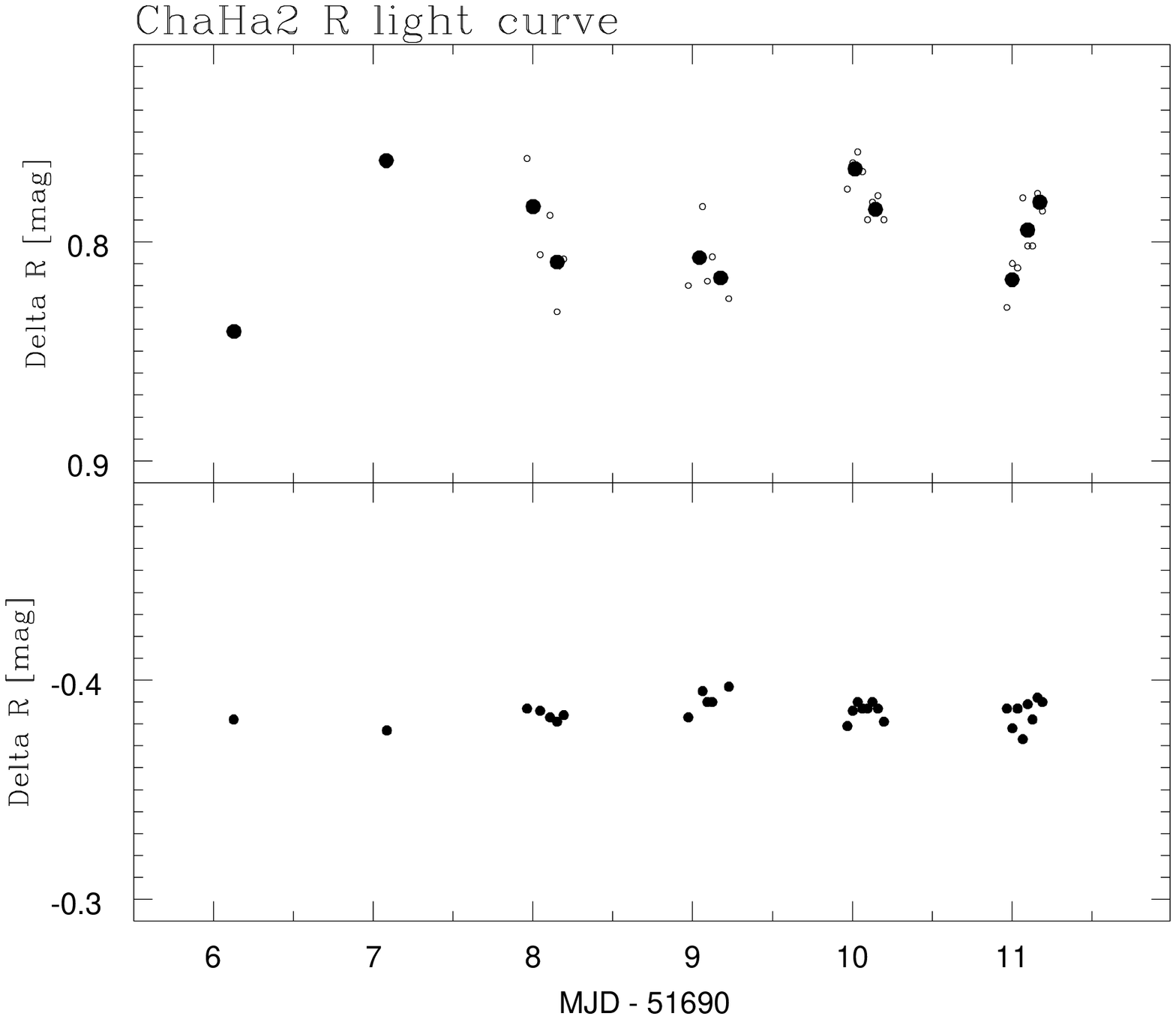}
\includegraphics[width=.5\textwidth]{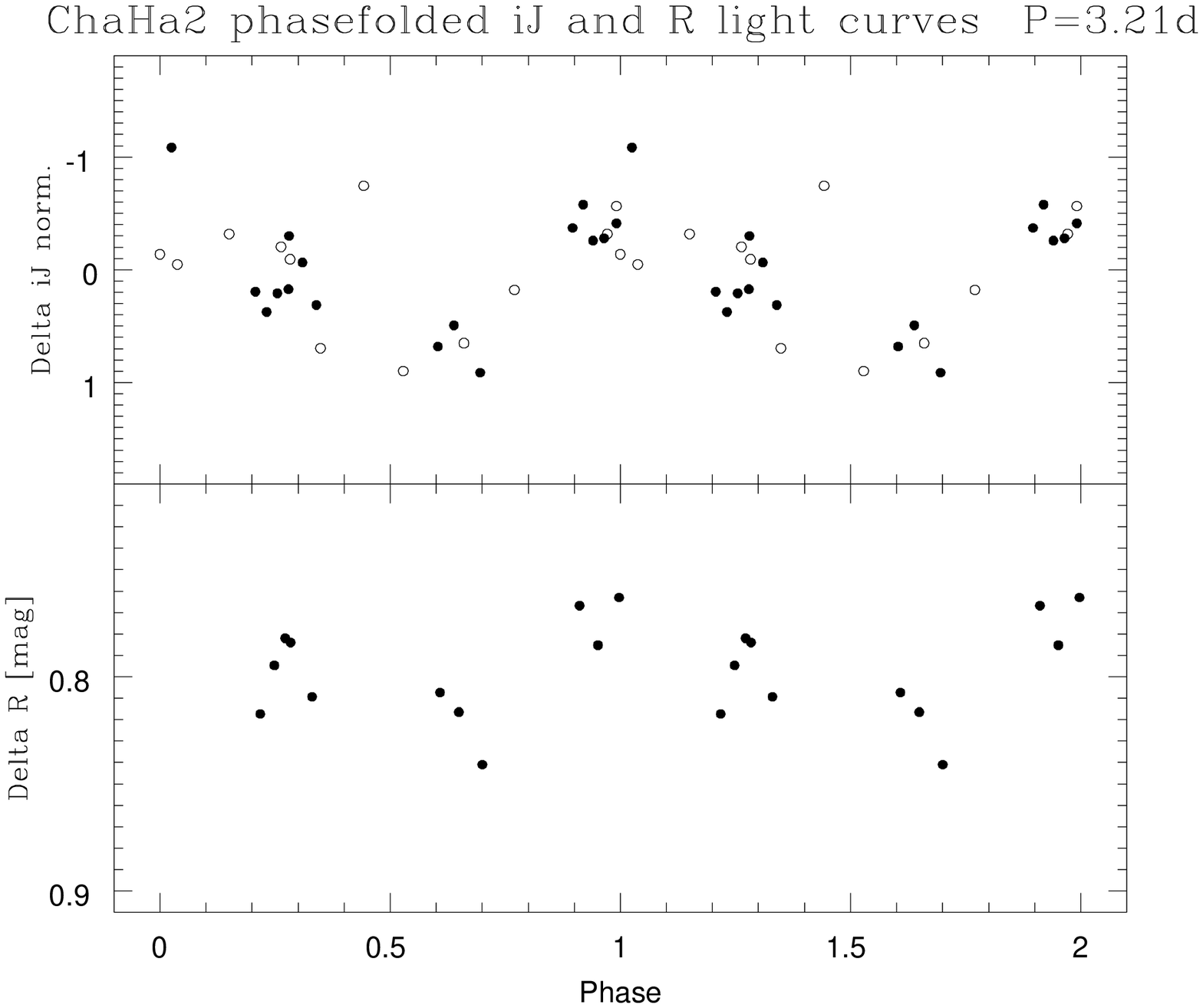}
\end{center}
\caption[]{\label{cha2_lc} Light curves 
of the brown dwarf candidate Cha\,H$\alpha$\,2.
Panels are as in Fig.\,\ref{b34_lc}.
Filled circles in the Gunn i and R light curves
(upper and middle panel) represent an average of the original
data points (open circles).
}
\end{figure}

In order to increase the S/N for the faint brown dwarf candidate
Cha\,H$\alpha$\,2 we binned adjacent data points in groups of about five.
Fig.\,\ref{cha2_lc} (top and middle panel) displays the original data
as well as the overplotted binned data.
The binned light curves have variability amplitudes of 0.05\,mag in Gunn i 
and 0.06\,mag in R.
We have found a minimum string length for a period of 2.8\,d 
for the Gunn i band data with a confidence level larger than 99\%. 
The light curves folded with this period 
support this result.
The data analysis yields also the two less significant periods 0.73\,d 
and 1.56\,d, which cannot be rejected at once by a look to the phased 
light curves.
However, they are related with the more significant 
2.8\,d period by Eq.\,\ref{alias} 
and are therefore supposed to be alias periods inherent to the data due 
to the sampling frequency of one day. 
%
The R band light curve of Cha\,H$\alpha$\,2 shows 
similar variations and trends as the Gunn i band light curve
(Fig.\,\ref{cha2_lc}, middle).
However, the period analysis of the R band data does not
confirm the 2.8\,d period on a high confidence level. 
It gives a minimum string length for 1.6\,d (the alias in the Gunn i band)
and only a second minimum for 2.8\,d.
The enhancement of the alias compared to the true period 
can be attributed to the smaller number of 
R band images as well as their lower S/N.

The period analysis for this target 
carried out on the Gunn i and J joined data set
gives 3.21$\pm$0.17 days as the best period. 
The Gunn i and R band data folded with this period show a smooth
curve (Fig.\,\ref{cha2_lc}, bottom).

We note that there are hints in HST images of Cha\,H$\alpha$\,2 
that it may be a close 0.2$^{\prime\prime}$ binary (Neuh\"auser et al. 2002)
unresolved in our DFOSC aperture photometry.
If this is confirmed,
the period found for Cha\,H$\alpha$\,2 is 
presumably the period of one of the
two components or a combination of both.

\subsection{Cha\,H$\alpha$\,3 ($\sim$ 0.06\,M$_{\odot}$)}
\label{chaha3}

\begin{figure}[h]
\begin{center}
\includegraphics[width=.5\textwidth]{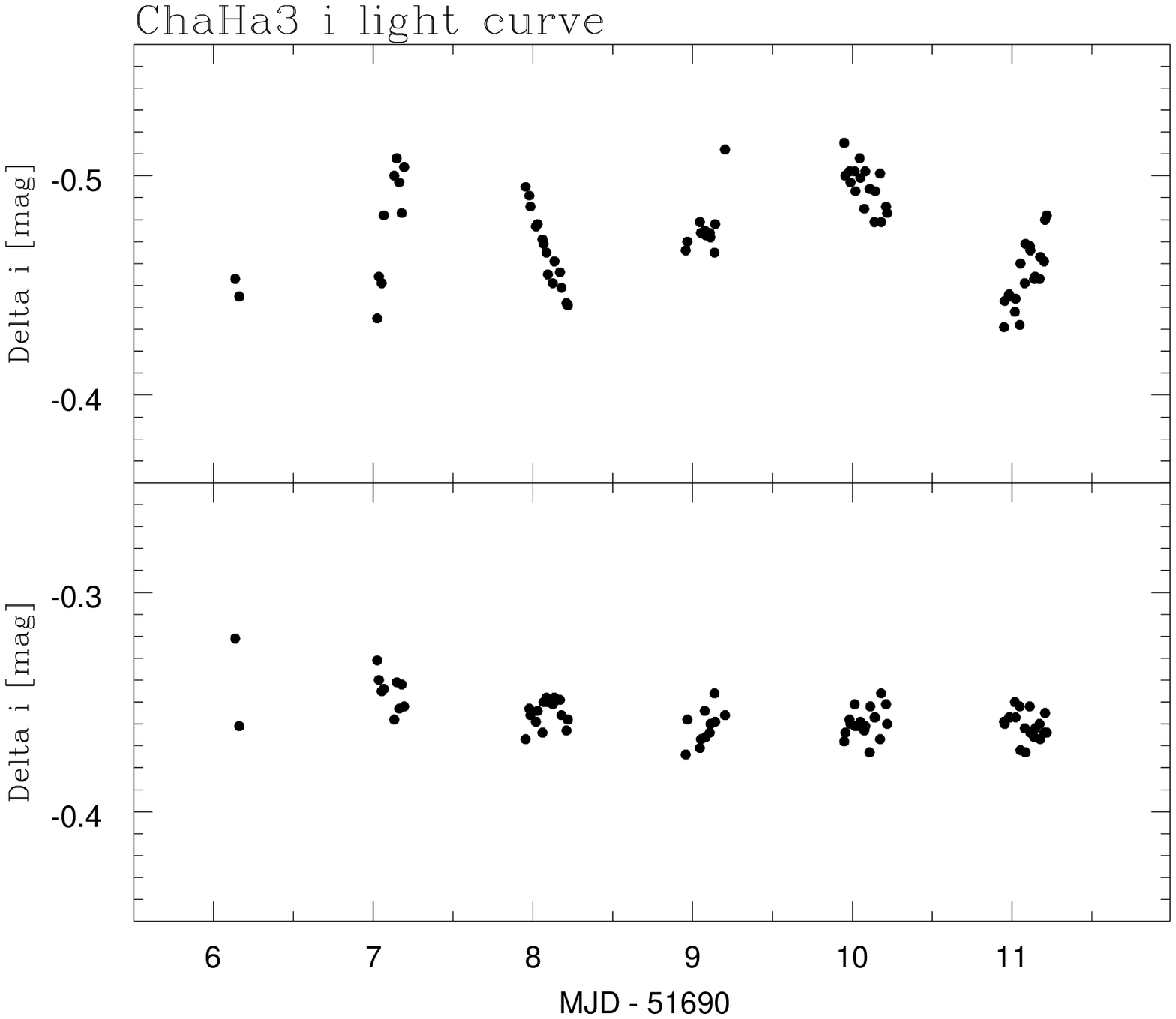}
\includegraphics[width=.5\textwidth]{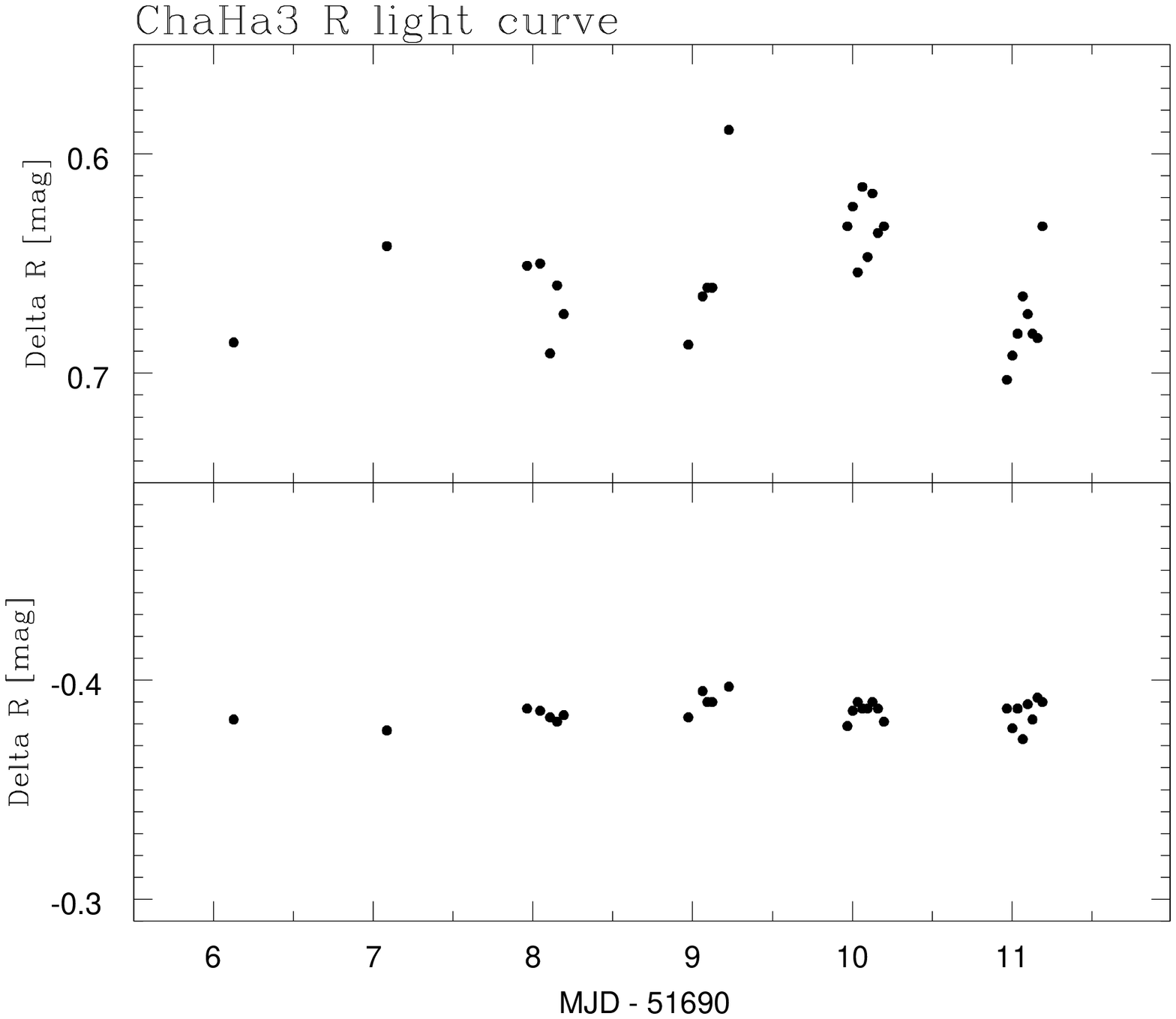}
\includegraphics[width=.5\textwidth]{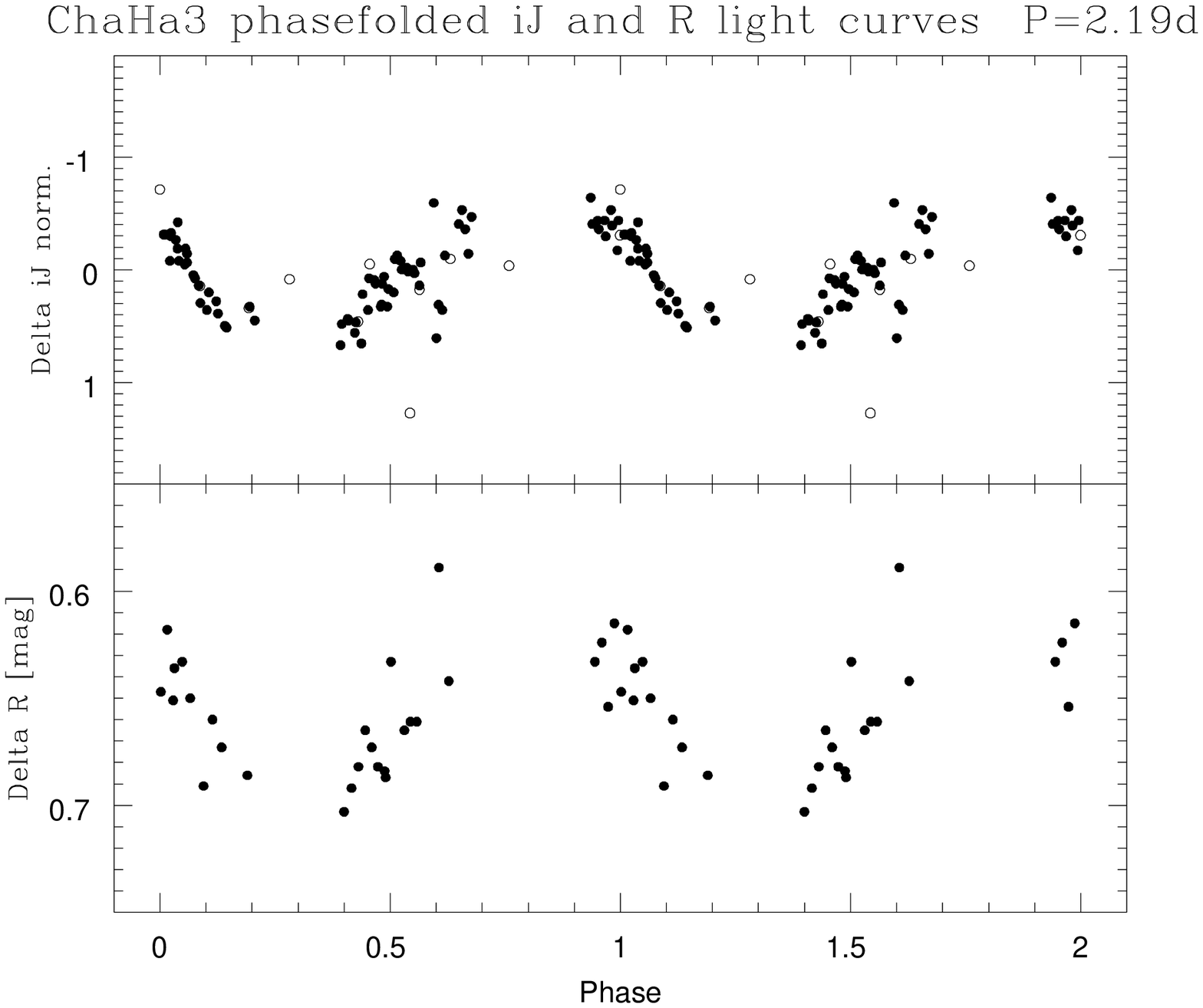}
\end{center}
\caption[]{\label{cha3_lc} Light curves 
of the brown dwarf candidate Cha\,H$\alpha$\,3.
Panels are as in Fig.\,\ref{b34_lc}.
}
\end{figure}

The brown dwarf candidate 
Cha\,H$\alpha$\,3 exhibits variations with a Gunn i amplitude of about 0.08\,mag 
and an R amplitude of about 0.09\,mag (Fig.\,\ref{cha3_lc}). 
The Gunn i band flux of this object seems to be modulated with a clear sine wave.
However, the first three data points in the second night
deviate from this behavior.
The period analysis without these three points yields
a highly significant period of 2.2\,d with a confidence level above 99.99\%.
This suggests that 2.2\,d is an intrinsic period to this object, although
strong reasons have not been found to reject the three deviant data points.
The R band data reflect the general trends
also seen in the Gunn i band but are more noisy. They support a period of 
2.2\,d but do not confirm it with high significance.
The R band data of nights 3 to 6 show a minimum string length
for a period of 0.06\,d but this can be rejected on the basis of 
a check of the data folded with this period. 
There is a second minimum at 2.2\,d and the R band data folded with 
2.2\,d look quite smooth. 
 
The period analysis carried out on the Gunn i and J joined data set
yields a period of 2.19\,d $\pm$ 0.09\,d for Cha\,H$\alpha$\,3,
after discarding one single outlying J data point.
The light curves of this object folded with the derived period 
also confirm the result and are displayed in Fig.\,\ref{cha3_lc} (bottom).
The outlying J data point was included again for the plot for completeness.
It is the lowest one in the joined iJ light curve at a phase of about 0.55.

\subsection{Cha\,H$\alpha$\,6 ($\sim$ 0.05\,M$_{\odot}$)}

\begin{figure}[h]
\begin{center}
\includegraphics[width=.5\textwidth]{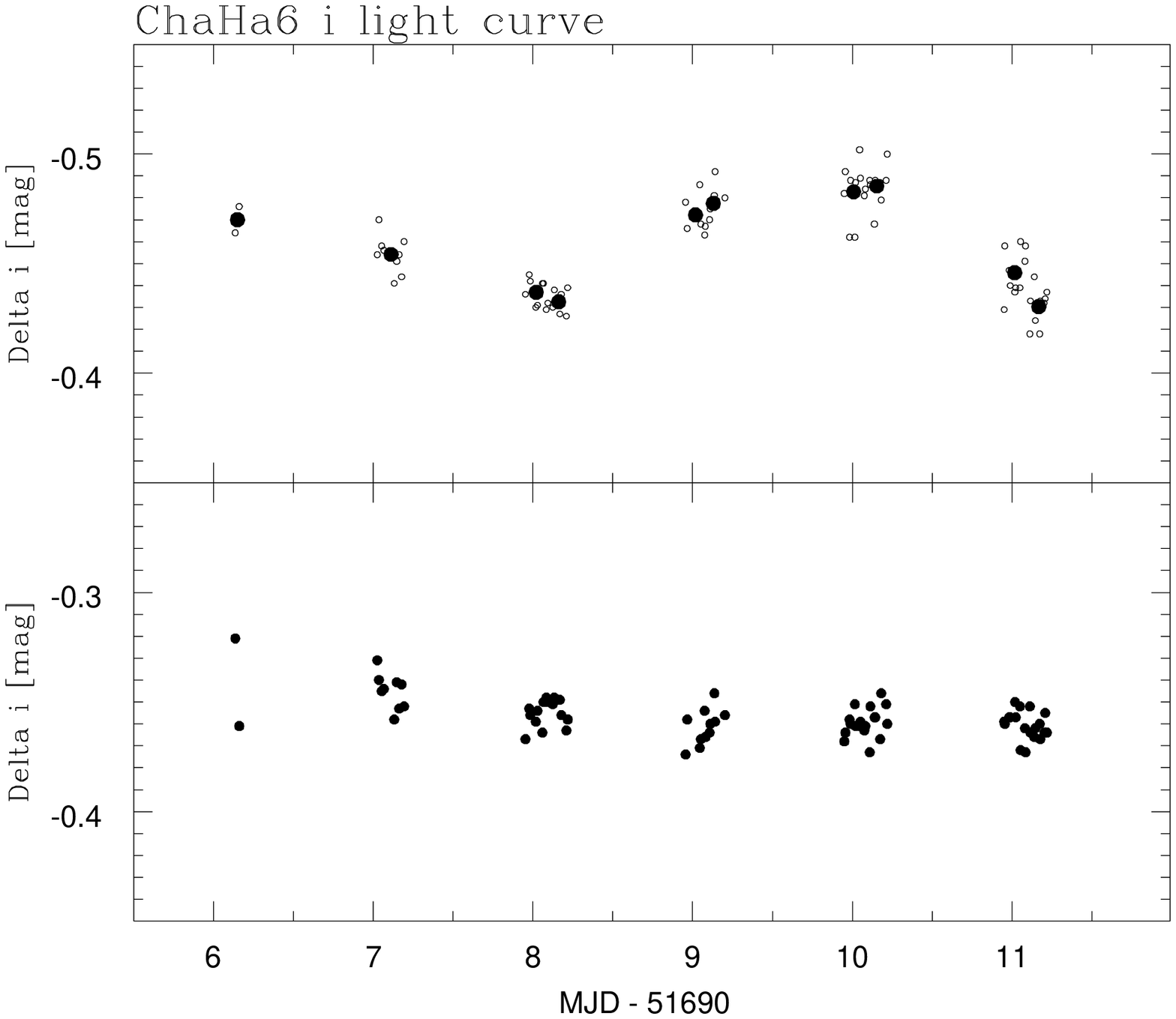}
\includegraphics[width=.5\textwidth]{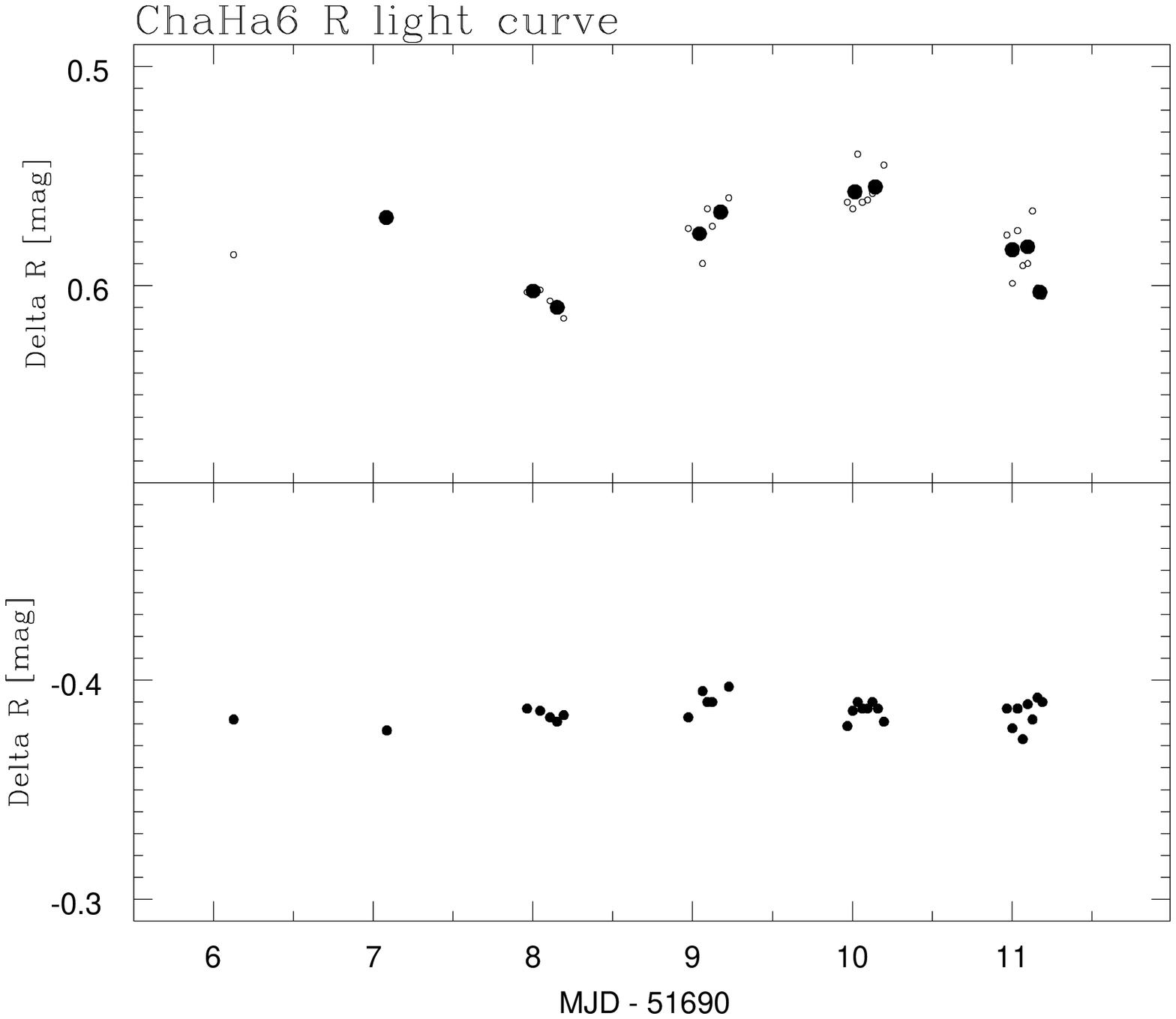}
\includegraphics[width=.5\textwidth]{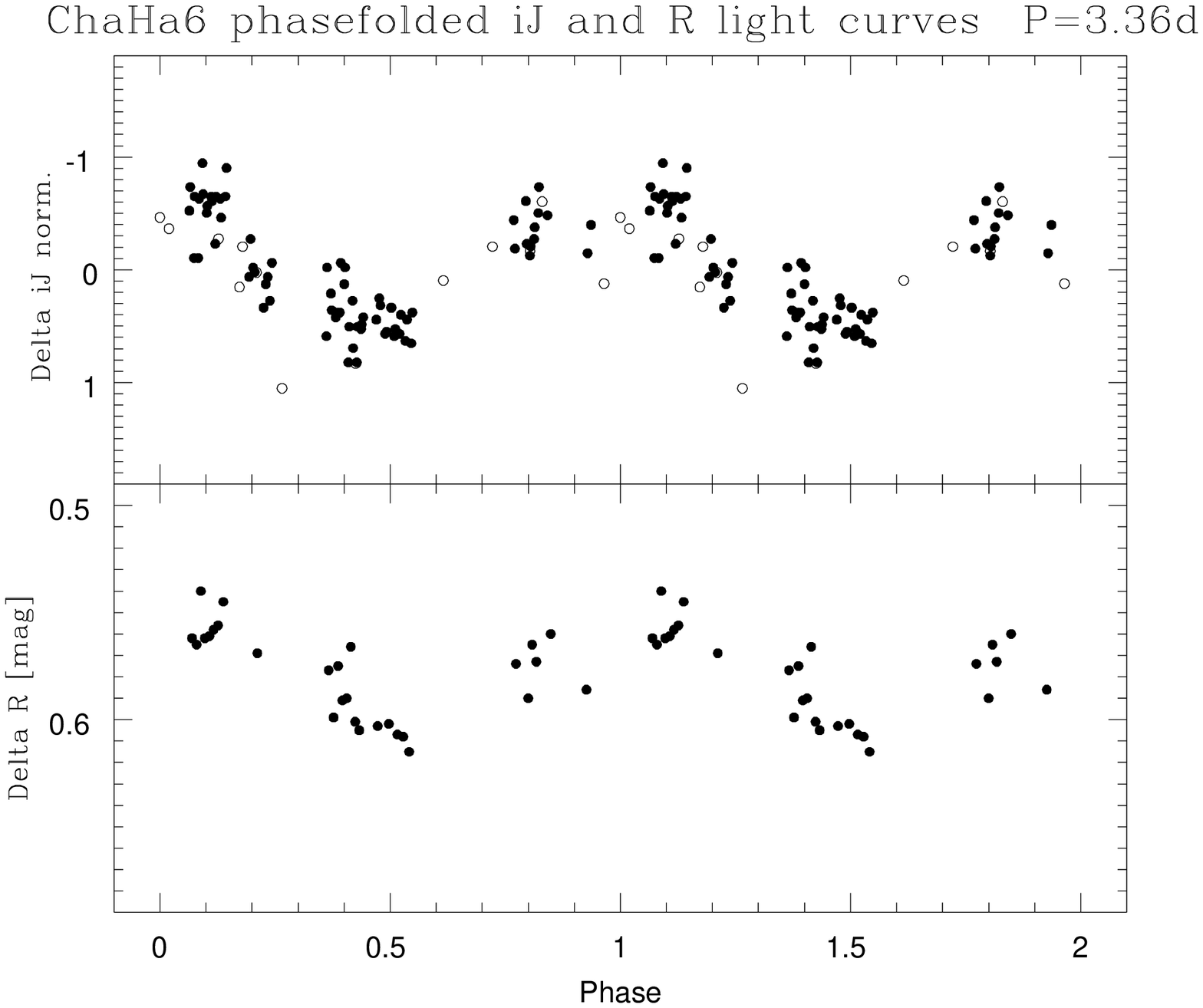}
\end{center}
\caption[]{\label{cha6_lc} Light curves of the brown dwarf
candidate Cha\,H$\alpha$\,6.
Panels are as in Fig.\,\ref{b34_lc}.
Filled circles in the Gunn i and R light curves
(upper and middle panel) represent an average of the original
data points (open circles).
}
\end{figure}

In order to increase the S/N of the brown dwarf candidate Cha\,H$\alpha$\,6,
we binned adjacent data points
in groups of two or three. See Fig.\,\ref{cha6_lc} (top and middle panel)
for the original
data and the overplotted binned data.
The binned data of the object shows variabilities with an amplitude of 
0.06\,mag in both bands.
The study of periodicities for the Gunn i band data
results in a minimum string length for a period of 3.49\,d
with a confidence level of 96\%.
There are two less significant minima at 0.77\,d and 1.43\,d,
which are related to the 3.49\,d period
by Eq.\,\ref{alias}. 
The analysis of the R band data is hampered by the small number of
data points. Particularly in the first night there is only
a single data point which has a large uncertainty due to the 
poor weather conditions. 
We ignored it for the further analysis of the 
R band data and the search for periods. 
We find a minimum string length for a period 
of 3.34\,d and two less significant minima for 0.78\,d and 
1.58\,d. The alias periods for 3.34\,d are 0.77\,d and 1.42\,d,
therefore 0.78 and 1.58 may be aliases. 

The period analysis carried out for the Gunn i and J joined data 
set reproduces exactly the period
that was obtained for the Gunn i and R band data, 3.36$\pm$0.19 days.
The confidence level is 99.99\%. 
The phase folded light curves of Cha\,H$\alpha$\,6 also 
confirms it as the best period (cf. Fig.\,\ref{cha6_lc} bottom). 

\section{Discussion}
\label{discussion}

The periods measured for Cha\,H$\alpha$\,2, Cha\,H$\alpha$\,3 and Cha\,H$\alpha$\,6 
are among the first photometric periods determined for very young 
substellar objects and the longest found for any brown dwarf so far. 
Furthermore, the objects have relatively large variability amplitudes compared
to hitherto monitored brown dwarfs.
Causes of these variations as well as constraints for the evolution of 
angular momentum in the substellar regime are discussed in the following sections.

\subsection{Rotational modulation due to spots}

We interpret the detected periodic photometric variations of the brown dwarf
candidates as well as of the T~Tauri stars as rotational modulation of the 
emitted flux due to surface spots 
based on the following arguments:

(1) The detected periods are consistent with rotational
velocities $v \sin i$ and are therefore likely rotational periods. 
See Sect.\,\ref{maxrotP} for more details.

(2) The monitored (R-i) colors of these objects show larger amplitudes 
for shorter wavelength in agreement with the expectations for spots. This
is true for all but Cha\,H$\alpha$\,6, which shows
the same amplitude in both the Gunn i and R band.  
Nevertheless, there may be slight differences between the 
amplitudes in the two bands, which are swallowed by the measurements 
uncertainties. The J band amplitudes based on the data by Carpenter et al. (2002)
for the same objects are always smaller or similar than
those observed in the Gunn i band also in agreement with spots.
Only for Cha\,H$\alpha$\,6, the J band amplitude is apparently
slightly larger than the Gunn i amplitude.
However, the fact that the 
J band data have larger photometric errors than the R and Gunn i data
could very well account for the larger J band amplitude for Cha\,H$\alpha$\,6.

(3) Young, pre-main sequence stars are known to have dominant surface 
features (e.g. Bouvier et al. 1993 and references therein), 
which are attributed to magnetic dynamo action.
Though the brown dwarfs as well as the T~Tauri stars 
studied in this paper
have very low masses, they are relatively hot due to their young age.
Thus, the atmospheric temperatures may be 
hot enough for a sufficient ionization fraction of the plasma in 
the atmospheres
to allow for magnetically induced surface features.
(It should be noted that the dynamo operating in fully convective objects,
like pre-main sequence stars, stars of lower mass than $\sim$0.4\,M$_{\odot}$
and brown dwarfs, is not yet completely understood.)

(4) The temperatures of these young objects are too hot 
(between 3030\,K for B\,34 and 2840\,K for Cha\,H$\alpha$\,3 and 6)
for significant dust condensation according to model atmosphere 
calculations 
(e.g. Tsuji, Ohnaka \& Aoki 1996a; Tsuji et al. 1996b; 
Allard et al. 1997; Burrows \& Sharp 1999). 
Therefore, non-uniform condensate-coverage that has been
suggested to explain the photometric variability detected for some 
L and M type old and hence cool (brown) dwarfs (Tinney \& Tolley 1999; 
Bailer-Jones \& Mundt 2001; Mart\'\i n et al. 2001; Clarke et al. 2002)
is an unlikely cause for the variabilities 
detected by us for the studied Cha\,I objects.

\subsection{Comparison with spectroscopic velocities $v \sin i$}
\label{maxrotP}

As mentioned above, the detected periods for Cha\,H$\alpha$\,2, 3, 6 
and B\,34 are consistent within the measurements uncertainties
with their projected rotational velocities $v \sin i$ 
of 12.8\,km\,s$^{-1}$, 21.0\,km\,s$^{-1}$, 13.0\,km\,s$^{-1}$ and 
15.2\,km\,s$^{-1}$, respectively (Joergens \& Guenther 2001). 
This indicates that they are likely rotational periods of the objects
(for CHXR\,78C no $v \sin i$ measurement is available).
Rotational periods $P_{v \sin i}$, which are upper limits of the true 
rotational periods
have been derived from radii and
$v \sin i$ values and are given in Table\,\ref{largetable}.
The radii have been determined based on luminosities and effective 
temperatures from Comer\'on et al. (1999, 2000) applying the 
Stefan-Boltzmann law.

It is noteworthy that the observed periods $P_{phot}$
are always -- with the exception of
Cha\,H$\alpha$\,2 - larger than the calculated periods $P_{v \sin i}$ 
but do still agree with 
them when the errors are taken into account, hinting at a systematic effect. 
A 1\,$\sigma$ error of $P_{v \sin i}$ with propagated 
errors of the luminosity, the effective temperature 
($\Delta \log L_\mathrm{bol}/L_{\odot} = 0.175$,
$\Delta T_\mathrm{eff} = 150\,K$, cf. Comer\'on et al. 2000)
and $v \sin i$ ($\Delta v \sin i$ =1--3\,km\,s$^{-1}$, 
cf. Joergens \& Guenther 2001)
is given in Table\,\ref{largetable}.
P$_{v \sin i}$ scales with $\sqrt{L_\mathrm{bol}}$, $T_\mathrm{eff}^{-2}$ 
and $v \sin i^{-1}$. Reasons for systematic deviations of
the estimated $P_{v \sin i}$ from the observed $P_\mathrm{phot}$ 
may be either that 
the luminosities (bolometric correction, distance) have been 
systematically underestimated;
the rotational velocities $v \sin i$ have been systematically overestimated;
the effective temperatures have been systematically overestimated
and/or the radii are systematically too small.

\subsection{Radii of brown dwarfs}
\label{radii}

The radii of the very young brown dwarfs and 
brown dwarf candidates Cha\,H$\alpha$\,1 to 12 
estimated from luminosities and effective temperatures are
ranging from 0.3\,R$_{\odot}$ for Cha\,H$\alpha$\,11 ($\sim$0.03\,M$_{\odot}$)
to 0.9\,R$_{\odot}$ for Cha\,H$\alpha$\,4 ($\sim$0.1\,M$_{\odot}$).
The uncertainties are of the order of 30\% due to propagated errors
in the luminosities and effective temperatures.
Their relatively large radii 
are fully consistent with the extremely young age of the objects
and the fact that they are still in a contracting stage.
Models of Chabrier \& Baraffe (1997) show that a brown dwarf 
with a mass of 0.06\,M$_{\odot}$ has a radius of $\sim$0.55\,R$_{\odot}$ at
3\,Myr and a radius of $\sim$0.45\,R$_{\odot}$ at 5\,Myr in their calculations.
Only after $\sim$500Myr, brown dwarfs have more or less shrunken to their 
final size at a radius of $\sim$\,0.1\,R$_{\odot}$
independent of their mass (Chabrier et al. 2000). 

Bailer-Jones \& Mundt (2001) studied the photometric 
variability of late-M and L dwarfs, among them several very young 
objects in $\sigma$\,Orionis. The authors
claim that none of the objects in their sample
has a radius larger than 0.2\,R$_{\odot}$. 
Based on this radius assumption 
and on spectroscopic velocities $v \sin i $ of 10-60\,km\,s$^{-1}$,
as found by Basri et al. (2000) for late-M and L dwarfs in the field, 
they infer expected rotational periods
for their targets of the order of 1 to 10 hours (see also Sect.\,\ref{evol}).
However, the radii for the $\sigma$\,Orionis objects in their sample,
which have an age of 1--5\,Myr
and masses between 0.02 and 0.12\,M$_{\odot}$,
are certainly larger than 0.2\,R$_{\odot}$.
This is not only shown by a comparison with theoretical models 
(e.g. Chabrier et al. 2000) but there is also
observational evidence for it from the study of the L1.5 dwarf S\,Ori\,47.
Estimations of its effective temperature 
and luminosity (Zapatero Osorio et al. 1999)
indicate that the radius even for this very low-mass brown dwarf
($\sim$0.015\,M$_{\odot}$)
is of the order of 0.35\,R$_{\odot}$. 
The S\,Ori objects studied by Bailer-Jones \& Mundt (2001) are of
significantly higher mass than S\,Ori\,47
and have therefore most certainly much larger radii than it.

\subsection{Evolution of angular momentum}
\label{evol}

In the current picture of angular momentum evolution, 
young solar-mass stars as well as brown dwarfs are supposed to undergo a spin up 
due to contraction on the Hayashi track. 
Interaction with an accretion disk can hold up the acceleration of the 
rotation by magnetic braking until the inner disk is dissipated; then the star
begins to spin up.
As the star ages on the main sequence, 
angular momentum loss through stellar winds spins down the star again. 
In contrast, at the very low masses of brown dwarfs, there is supposed to
be no braking due to winds, which would explain the observed
(very) fast rotation of old brown dwarfs, which
have $v \sin i$ values ranging from 
5\,km\,s$^{-1}$ up to 
60\,km\,s$^{-1}$ (Basri \& Marcy 1995; Mart\'\i n et al 1997; 
Tinney \& Reid 1998; Basri et al. 2000).
Furthermore, brown dwarfs with ages between about 36\,Myr and 1\,Gyr seem 
to have rotational periods shorter than one day 
(see Sect.\,\ref{hitherto} for references).
The lower age limit at about 36\,Myr is set by rotational periods 
determined for brown dwarf candidates in the young cluster IC\,4665 
(Eisl\"offel \& Scholz 2001).

However, the periods that we have measured for three
brown dwarf candidates in Cha\,I range between 2 and 
3 days and are therefore
larger than any period reported up to now for such low-mass objects.
The relatively long rotational periods and moderately fast
rotational velocities of the brown dwarf candidates in Cha\,I
are explained naturally within the current picture of angular momentum 
evolution by 
the fact that they are still in an early contracting stage. They
may have furthermore
suffered until very recently a braking due to their interaction
with an accretion disk. 

The comparison of our rotation periods at 1--5\,Myr 
with those in the literature at $>$\,36\,Myr,
gives first indications that
most of the acceleration of brown dwarfs takes place 
in the first 30\,Myr or less of their lifetime.
It is known that Cha\,H$\alpha$\,2 and 6 have optically thick disks 
(Comer\'on et al. 2000), therefore magnetic braking due to 
interactions with the disk may play a role for them.
This is suggested by the fact, that among 
the three brown dwarf candidates with determined periods, the one without
a detected disk, Cha\,H$\alpha$\,3, has the shortest period.
If the interaction with the disk is responsible for the braking,
our results and the one from Eisl\"offel \& Scholz (2001) indicate that 
brown dwarf inner disks dissipate between 1--5\,Myr and 36\,Myr.
These limits for the time scale of disk dissipation
for brown dwarfs are not inconsistent with that for T~Tauri stars, which are
known to dissolve their inner disks within about the first 10\,Myr 
(e.g. Calvet et al. 2000). 

Bailer-Jones \& Mundt (2001) report the detection of photometric periods for
the two very young (1--5\,Myr) M6.5 dwarfs
S\,Ori\,31 and S\,Ori\,33 of less than 9\,h (cf. Table\,\ref{literature})
and suggest that these 
might be rotational periods of the objects.
However, as described in Sect.\,\ref{radii}, a radius of $<$0.2\,R$_{\odot}$
is unreliable for the young S\,Ori objects;
a value of the order of half to one solar radius is more consistent
(e.g. Chabrier et al. 2000).
If the periods of less than 9\,h for S\,Ori\,31 and S\,Ori\,33 are 
confirmed as rotational periods, their rotational velocities would be
of the order of 100\,km\,s$^{-1}$,
assuming for example a radius of 0.6\,R$_{\odot}$.
This would mean that these very low-mass stars rotate at a much higher
speed than the brown dwarfs and brown dwarf candidates in Cha\,I, 
which have spectroscopic velocities
$v \sin i$ of 9--28\,km\,s$^{-1}$ (Joergens \& Guenther 2001)
and rotational periods of 2 to 3 days as shown in this paper.
A possible explanation for a much faster rotation of the S\,Ori objects
compared to the brown dwarfs in Cha\,I despite the similar age,
would be a significant difference in their disk
life-times, at least if magnetic braking due to a circumstellar 
disk is an important process for them.
The dissipation of circumstellar material in $\sigma$ Orionis
could be enhanced due to strong winds from the hot OB stars in this region
in comparison with the Cha\,I region without such a hot star.
Therefore disk life-times could be much shorter in $\sigma$ Orionis
leading to an early spin up and, thus, to significantly faster rotation
of the S\,Ori objects than of the brown dwarfs and very low-mass stars in Cha\,I.

A measurement of $v \sin i $ of S\,Ori\,31 and S\,Ori\,33
would be useful to confirm their proposed very fast rotation.
However, high-resolution spectroscopy for these faint objects is 
challenging even with large telescopes.

\section{Summary}
\label{concl}
We have determined photometric periods in the Gunn i and R band for the 
three M6.5--M7 type brown dwarf candidates 
Cha\,H$\alpha$\,2, Cha\,H$\alpha$\,3 and Cha\,H$\alpha$\,6 
of 3.2\,d, 2.2\,d and 3.4\,d, respectively.
	These are the longest photometric periods found for any brown 
	dwarf so far. If interpreted as rotationally induced they 
	correspond to moderately fast rotational velocities, which is fully 
	consistent with their $v \sin i$ values 
	(Joergens \& Guenther 2001) and their relatively large radii.
	Furthermore, we have detected periods for the two M5--M5.5 type
	very low-mass stars B\,34 and CHXR\,78C of 4.8\,d and 4.3\,d.
For the determination of the periods, we took in addition to our
optical Gunn i and R photometry also 
J band data obtained by Carpenter et al. (2002) into account. 
Our Gunn i and R band data are sampled with a relatively high frequency
($\sim$0.5\,h) but have a relatively short time base (5.5\,d), 
whereas the J band
data consist of 10--24 data points spread over 41 nights
immediately before our observations
and are therefore ideally complementing our data
and increasing the total time coverage to 55--59 days.

The observed periodic variations of the brown dwarf candidates as 
well as of the T~Tauri stars are interpreted as modulation of the 
flux at the rotation period due to magnetically driven surface 
features on the basis of a consistency with $v \sin i$ values as 
well as (R-i) color variations typical for spots. 
Furthermore, the temperatures even for the brown dwarfs in the sample 
are relatively high ($>$\,2800\,K) because the objects are very 
young. Therefore,
the atmospheric gas should be sufficiently ionized for the 
formation of spots on one hand and the temperatures are
too high for significant dust condensation and hence variabilities due
to clouds on the other hand.
These first indications for surface spots on very young brown dwarfs
support the overall picture of magnetic activity of brown dwarfs, which is 
emerging in the last years: old (cool) brown dwarfs tend towards an absence
of persistent activity, whereas very young brown dwarfs
seem to have more in common with T~Tauri stars as they 
are relatively active in terms of X-ray emission 
(e.g. Neuh\"auser et al. 1999) and H$\alpha$ emission. 

Estimation of the radii for the bona fide and candidate brown dwarfs
Cha\,H$\alpha$\,1 to 12 show that they are relatively large 
(0.3--0.9\,R$_{\odot}$) for such low-mass objects. 
They are fully 
consistent with theoretical calculations of the contraction time 
scales of such very young brown dwarfs and objects near the hydrogen
burning mass limit.

The periods measured by us provide valuable data points in an as yet, in terms
of rotational characteristic, almost unexplored region of the age-mass diagram.
A comparison of the determined rotational periods 
at the age of a few million years with rotational properties
of older brown dwarfs ($>$36\,Myr, Eisl\"offel \& Scholz 2001)
shows that most of the acceleration of brown dwarfs
during the contraction phase takes place within the first 30\,Myr or less
and suggests that the disk dissipation for brown dwarfs occurs between
1--5\,Myr and 36\,Myr.

However, a larger sample of brown dwarfs at various ages and in particular 
at a few million years
with known rotation rates
is much needed to further constrain the substellar 
angular momentum evolution as well as to allow for statistics of
rotation rates in connection with activity indicators.




\acknowledgments
We like to acknowledge helpful 
discussions on the topic of this paper with 
G. Wuchterl, B. Stelzer and C. Broeg. 
Furthermore, we thank our referee, C. Bailer-Jones,
for valuable comments, which helped to improve the paper.
We are also grateful to E. Guenther for 
contributions to the photometric monitoring campaign,
R. Garrido for his help concerning the uncertainties
of the period determination and
to the ESO staff at La Silla for their support during the DFOSC observations.
      VJ acknowledges grant from the Deutsche Forschungsgemeinschaft
      (Schwerpunktprogramm `Physics of star formation'). 
      MF was partially supported by the spanish grant PB97-1438-C02-02.	
      RN acknowledges
      financial support from the BMBF through DLR grant 50OR\,0003.






\clearpage



\clearpage 

\clearpage

\clearpage

\clearpage

\clearpage

\begin{table*}[t]
\caption{
\label{literature}
Photometric periods of brown dwarfs from the literature. 
}
\begin{tabular}{lccllll}
\tableline
\tableline
Object & Age & Mass          & SpT & Period & Amplitude & ref \\
       &     & [M$_{\odot}$] &     &        &  [mag]    &     \\ 

\tableline

S\,Ori\,31         & 1-5\,Myr  &         & (M6.5)  & 7.5\,h         & 0.012 & 1\\

S\,Ori\,33         & 1-5\,Myr  &         & (M6.5)  & 8.6\,h, 6.5\,h & 0.010 & 1\\  

CFHT-PL\,8         & 100-120\,Myr & 0.08 & $\dots$ & 9.6\,h         & 0.028 & 2 \\

LP\,944-20         & 475-650\,Myr & 0.06 & M9      & $\dots$ & 0.04$^{\ast}$ & 3 \\

BRI\,0021-0214     & $\geq$1\,Gyr && M9.5    & 19.2\,h, 4.8\,h & 0.018, 0.007 & 4 \\

2M1334             & $\geq$1\,Gyr && L1.5    & 2.7\,h, 6.3\,h, 1.0\,h & 0.020 & 1 \\

Kelu-1             & $\geq$1\,Gyr & $<$0.07 & L2 & 1.8\,h  & 0.012$^{\ast}$ & 5 \\

2M1146             & $\geq$1\,Gyr &         & L3      & 5.1\,h  & 0.015  & 1 \\

SDSS\,0539         & $\geq$1\,Gyr &         & L5      & 13.3\,h & 0.011 & 1 \\

2MASS\,0746+20AB   & $\geq$1\,Gyr &         & L0.5    & 31.0\,h & 0.010 & 6 \\

2MASS\,1300+19     & $\geq$1\,Gyr &         & L1      & 238.\,h & 0.015 & 6 \\

\tableline
\end{tabular}
\tablecomments{
S\,Ori\,31, S\,Ori\,33 and CFHT-PL\,8 are eventually stars but have been included 
because of their closeness to the hydrogen burning mass limit. Masses are 
estimates. Ages of the field brown dwarfs BRI\,0021-0214 to 2MASS1300 are 
unknown but supposed to be of the order of 1\,Gyr or more.
The spectral types of S\,Ori\,31 and S\,Ori\,33 are estimates. 
Amplitudes are measured in 
broad band I filters, except for the ones marked with an asterisk, these 
refer to narrow band observations (cf. references for details).  
Some object names are acronyms as given by the authors; their full names are:
S\,Ori\,31 = S\,Ori\,J053820.8-024613; 
S\,Ori\,33 = S\,Ori\,J053657.9-023522;
CFHT-PL\,8 = VLC\,J0342268+245021;
2M1334 = 2MASSW\,J1334062+194034;
2M1146 = 2MASSW\,J1146345+223053;
SDSS\,0539 = SDSSp\,J053951.99-005902.0;
2MASS\,0746+20AB = 2MASSI\,J0746425+200032;
2MASS\,1300+19 = 2MASSW\,J1300425+191235.
}
\tablerefs{
(1) Bailer-Jones \& Mundt 2001;
(2) Terndrup et al. 1999;
(3) Tinney \& Tolley 1999;
(4) Mart\'\i n et al. 2001;
(5) Clarke et al. 2002;
(6) Gelino et al. 2002.
}
\end{table*}

\clearpage

\begin{table*}[t]
\caption{
\label{largetable} 
Rotational periods and photometric amplitudes 
for brown dwarfs and very low-mass stars in Cha\,I
}
\begin{tabular}{l|cccc|cccc|ccc}
\tableline
\tableline
Object  & \multicolumn{4}{c|}{Characteristics of the targets} & \multicolumn{4}{c|}{Periods} & \multicolumn{3}{c}{Phot. amplitude}\\
	& SpT & $M_{\ast}$ & I & $R_{\ast}$ & $P_\mathrm{phot}$ & $\Delta P_\mathrm{phot}$ &
	$P_{v \sin i}$ & $\Delta P_{v \sin i}$ & $\Delta$\,R\, & 
	$\Delta$\,i\, & $\Delta$\,J \\
	&     & [M$_{\odot}$] & [mag] & [R$_{\odot}$] & [d] & [d] & [d] & [d] &
	[mag] & [mag] & [mag] \\
\tableline
B\,34               & M5   & 0.12 & 14.3 & 0.93 & 4.75 & 0.38 & 3.1 & $^{+1.7}_{-1.0}$ & 0.18 & 0.14 & 0.13 \\
CHXR\,78C           & M5.5 & 0.09 & 14.8 & 0.94 & 4.25 & 0.31 & $\dots$ & $\dots$      & 0.10 & 0.07 & 0.06 \\
Cha\,H$\alpha$\,2   & M6.5 & 0.07 & 15.1 & 0.73 & 3.21 & 0.17 & 2.9 & $^{+1.4}_{-1.0}$ & 0.06 & 0.05 & 0.05 \\
Cha\,H$\alpha$\,3   & M7   & 0.06 & 14.9 & 0.77 & 2.19 & 0.09 & 1.9 & $^{+0.8}_{-0.6}$ 
& 0.09  & 0.08 & 0.08 \\
Cha\,H$\alpha$\,6   & M7   & 0.05 & 15.1 & 0.68 & 3.36 & 0.19 & 2.6 & $^{+2.0}_{-1.0}$ & 0.06 & 0.06 & 0.10 \\
\tableline
CHXR\,73            & M4.5 & 0.15 & 15.6 & 1.46 & $\dots$ & $\dots$ & $\dots$ & $\dots$ & $<$0.03 & $<$0.05 
& $\dots$\\ 
Cha\,H$\alpha$\,4   & M6   & 0.1  & 14.3 & 0.89 & $\dots$ & $\dots$ & $\dots$ & $\dots$ & 
$<$0.03 & $<$0.03 & $\dots$ \\
Cha\,H$\alpha$\,5   & M6   & 0.1  & 14.7 & 0.83 & $\dots$ & $\dots$ & $\dots$ & $\dots$ & 
$<$0.04 & $<$0.02 & $\dots$\\
Cha\,H$\alpha$\,8   & M6.5 & 0.07 & 15.5 & 0.59 & $\dots$ & $\dots$ & $\dots$ & $\dots$
& $<$0.02 & $<$0.04 & $\dots$\\
Cha\,H$\alpha$\,12  & M7   & 0.05 & 15.6 & 0.66 & $\dots$ & $\dots$ & $\dots$ & $\dots$ & 
$<$0.04 & $<$0.03 & $\dots$\\
\tableline
\end{tabular}
\tablecomments{
Spectral types (SpT), masses ($M_{\ast}$) and I band magnitudes are 
from Comer\'on et al. (1999, 2000). Masses are estimates based on Baraffe et al.
(1998) tracks for the Cha\,H$\alpha$~objects and on Burrows et al. (1997) 
tracks for the T~Tauri stars, respectively. 
Radii ($R_{\ast}$) are estimated based on luminosities and effective 
temperatures from Comer\'on et al. (1999, 2000);
their errors are of the order of 30\%.
Rotational periods ($P_\mathrm{phot}$) and photometric peak-to-peak 
amplitudes ($\Delta$\,R, $\Delta$\,i, $\Delta$\,J) are derived from our Gunn i and R 
band light curves taking 
into account J band data from Carpenter et al. (2002).
$\Delta P_\mathrm{phot}$ is an upper limit for the error of the derived 
periods based on Nyquist's frequency.
$P_{v\sin i}$ denotes rotational periods derived from 
radii
and $v \sin i$ measurements (Joergens \& Guenther 2001). 
$\Delta P_{v \sin i}$ is a 1\,$\sigma$ error. 
Note that $\Delta$\,i refers to our Gunn i photometry, whereas the I band 
magnitudes by Comer\'on et al. are obtained in the Cousins I filter.
For CHXR\,78C no $v \sin i$ measurements are available.
See the text for more details.
}
\end{table*}


\end{document}